\documentclass[aps,prd,twocolumn,nofootinbib,groupedaddress, 10pt]{revtex4-2}
\usepackage[table,svgnames,dvipsnames]{xcolor}
\usepackage[colorlinks=true,citecolor=blue,linkcolor=blue,urlcolor=blue, backref=false,pdfborder={0 0 0}]{hyperref}
\usepackage[normalem]{ulem}
\usepackage{aas_macros}
\usepackage{amsmath,amssymb}
\usepackage{graphicx}
\usepackage{dcolumn}
\usepackage{bm}
\usepackage[mathlines]{lineno}
\usepackage{orcidlink}
\usepackage{cases}
\usepackage{booktabs}
\usepackage{comment}
\usepackage{multirow}
\usepackage{siunitx}
\usepackage{booktabs}
\usepackage{float}
\usepackage{geometry}
\newcommand{\Mpch}{$h^{-1}\,\mbox{Mpc}$}
\newcommand{\Gpch}{$h^{-1}\,\mbox{Gpc}$}

\newcommand{\Msunh}{$h^{-1}\,\mbox{M}_\odot$}

\newcommand{\Om}{$\Omega_{\rm m}$}
\newcommand{\Ob}{$\Omega_{\rm b}$}

\newcommand{\Mnu}{$M_\nu$}

\newcommand{\pairvel}{$v_{12}(r,a)$}
\newcommand{\lcdm}{$\mathrm{\Lambda CDM}$}
\newcommand{\vot}{$v_{12}$}  

\newcommand{\elephant}{\textsc{ELEPHANT}}
\newcommand{\quijote}{\textsc{QUIJOTE}}
\newcommand{\mdr}{\textsc{MDR1}}
\newcommand{\unitsim}{\textsc{UNIT}}

\newcommand{\refeq}[1]{Eq.~(\ref{eq:#1})}

\newcommand{\code}[1]{{\texttt{#1}}}
\renewcommand{\emph}[1]{\textit{#1}}

\begin{document}
 \preprint{arxiv}

\title{Parameter sensitivity of cosmic pairwise velocities in the non-linear regime of structure formation}
\author{Jorge Enrique Garc\'ia-Farieta\orcidlink{0000-0001-6667-5471}}\email{jegarciaf@unbosque.edu.co}
\affiliation{Laboratorio de Inteligencia Artificial (SavIA Lab), Grupo Signos, Departamento de Matemáticas, Universidad El Bosque, Bogotá, Colombia}
\author{H\'ector J. Hort\'ua\orcidlink{0000-0002-3396-2404}}
\affiliation{Laboratorio de Inteligencia Artificial (SavIA Lab), Grupo Signos, Departamento de Matemáticas, Universidad El Bosque, Bogotá, Colombia\\
Instituto de Neurociencias, Universidad El Bosque, Bogotá, Colombia
}

\date{\today}

\begin{abstract}
The peculiar velocities of dark matter tracers drive the growth of cosmic structures, providing a sensitive test of cosmological models and strengthening constraints on the nature of dark energy. In this work, we investigate the mean pairwise velocities, $v_{12}$, of dark matter tracers as a cosmological probe in the non-linear regime of cosmic structure formation. Using N-body dark matter-only simulations, we measure $v_{12}$ for pair separations up to $50$ \Mpch\ and model it by solving the pair conservation equation for a self-gravitating particle system, along with various prescriptions of the nonlinear matter power spectrum. We quantified the sensitivity of \vot\ to variations in key cosmological parameters such as \Om, $\sigma_8$, $h$, $M_\nu$, and $w$. Our parameter inference analysis using MCMC shows sub-11\% agreement with simulation data, with notable degeneracies, particularly between $\Omega_m$ and $\sigma_8$. We further compute the stable clustering crossing scale across redshifts $z=0$, $0.5$, and $1$, assessing its dependence on cosmology. Among the tested power spectrum modeling approaches, we find that the \code{CSSTEmu} emulator provides the most accurate predictions, with deviations below 5\% for $r > 10$ \Mpch\ at $z=0.5$. Our results are validated using independent simulation suites, demonstrating that our framework offers a robust method for extracting cosmological constraints from upcoming peculiar velocity data.
\end{abstract}

\maketitle

\section{\label{sec:intro}Introduction}
The large-scale motions of galaxies, known as peculiar velocities, are emerging as a powerful cosmological probe due to their sensitivity to gravitational interactions with the surrounding matter field \cite{2016MNRAS.463.4083A, 2012ApJ...751L..30H, 2014MNRAS.445.4267K}. These motions reflect the cumulative effects of structure growth and gravitational infall, offering a direct link to the underlying density field and its evolution. Understanding the gravitational clustering of galaxies remains a central goal of modern cosmology, particularly in light of the persistent uncertainties surrounding the physical origin of cosmic acceleration within the standard \lcdm\ framework \cite{2013PhR...530...87W}. Similarly, small-scale clustering has become an important probe of cosmology in light of the growing tension between cosmological parameters inferred from the cosmic microwave background and those derived from low-redshift probes \cite{2017MNRAS.465.1454H,2020A&A...641A...6P,2022ApJ...934L...7R}. These discrepancies, especially in the inferred matter density, \Om, and the amplitude of matter fluctuations, $\sigma_8$, highlight the need for accurate modeling of the non-linear regime of structure formation. Analytical mode\-ling of the density perturbations has advanced considerably in the quasi-linear and non-linear regimes, employing a variety of approaches to predict the non-linear matter power spectrum such as standard and regularized perturbation theory \cite{2012JCAP...11..029G,2010PhRvD..82f3522T,2004PhRvD..70h3007S,2002PhR...367....1B, 2014ascl.soft04012T} as well as the updated Effective Field Theory of Large Scale Structure \cite{2012JCAP...07..051B,2012JHEP...09..082C}, fitting formulae like \code{HALOFIT} and its subsequent refinements \cite{2003MNRAS.341.1311S,2012ApJ...761..152T,2015MNRAS.454.1958M}, and simulation-calibrated emulators \cite{2014ApJ...780..111H,2019PhRvD.100l3540W,2019MNRAS.484.5509E,2021MNRAS.507.5869A}. 

The precision of current and upcoming cosmological surveys, such as the Dark Energy Spectroscopic Instrument (DESI) collaboration \cite{2016arXiv161100036D, 2025JCAP...02..021A}, EUCLID \cite{2011arXiv1110.3193L, 2022A&A...657A..91E}, Javalambre Physics of the Accelerating Universe Astrophysical Survey (J-PAS) \cite{2014arXiv1403.5237B}, Subaru Prime Focus Spectrograph (PFS), The China Space Station Telescope (CSST) \cite{2025SCPMA..6880402G}, and the 4-metre Multi-Object Spectroscopic Telescope (4MOST-CRS) \cite{2025arXiv250807311V}, makes extracting cosmological information from small scales is both timely and essential. The advances in redshift surveys, combined with improved and targeted measurements of peculiar velocities, have progressed to the point where mapping the peculiar velocity field is becoming achievable. This mapping will provide datasets capable of delivering high-precision measurements of the growth rate of cosmic structure. These measurements will significantly tighten constraints on models of gravity and dark energy, providing information that is complementary to that obtained from density field statistics alone \cite{2023MNRAS.518.2436T, 2024MNRAS.532.3972L}. 

Velocity statistics are emerging as a rapidly developing area of research, offering new avenues to extract cosmological information beyond density-only based methods. In particular, observables such as the mean pairwise peculiar velocity, \vot$(r)$, encode both cosmological expansion (Hubble flow) and gravitationally induced bulk motions, enriching the information content available for cosmological inference.

The pairwise velocity statistic \vot\ has proven to be a powerful tool for constraining key cosmological parameters. The seminal work by Ferreira et al. \cite{1999ApJ...515L...1F} established early estimators of pairwise velocity from redshift surveys and N-body simulations, which produced some of the first measurements of the matter density parameter \Om. Building on this foundation, \vot\ has also been applied to measure the Hubble constant \cite{2025ApJ...978L...6Z} and to detect the kinematic Sunyaev-Zel'dovich effect \cite{2016MNRAS.461.3172S}. Complementary efforts have focused on quantify the statistical accuracy of pairwise velocity estimators, see Ref. \cite{2023MNRAS.525.1039M} and to describe the scale and redshift variation of velocity and density distributions of self-gravitating collisionless systems \cite{2022arXiv220206515X}. Furthermore, pairwise velocities have been used to probe deviations from the standard \lcdm\ model; for example, recent works such as Refs. \cite{2024PhRvD.109l3528J} have applied \vot\ modeling to modified gravity cosmologies, while Ref. \cite{2017MNRAS.467.1386B} studied pairwise velocities within the ``Running FLRW'' cosmological model, which incorporates quantum field theory effects in curved spacetime. More recently, \vot\ has found applications in neutrino cosmology, aiding in measurements of neutrino mass and asymmetry \cite{2024MNRAS.529..360Z}.

In this paper, we investigate the pairwise velocities of dark matter tracers within the \lcdm\ framework, accounting for the effects of massive neutrinos and minimal variations in the dark energy equation of state. We focus on accurately modeling the mean pairwise velocity \vot, particularly on small scales, by solving the full pair-conservation equation combined with a range of non-linear matter power spectrum prescriptions, including emulator-based models and fitting functions. Using multiple independent sets of dark-matter-only N-body simulations, we measure how variations in key cosmological parameters, such as \Om, $\sigma_8$, $h$, the summed neutrino mass $M_\nu$, and the dark energy equation-of-state parameter $w$, impact \vot. This allows us to quantify the sensitivity of \vot\ to these parameters and identify which ones can be robustly extracted from the cosmological data. We validate our modeling approach against independent simulation suites, achieving a few percent accuracy in reproducing the simulation results. Furthermore, we perform a Markov Chain Monte Carlo (MCMC) analysis to extract cosmological constraints from \vot\ measurements and evaluate the robustness and precision of these constraints given the simulated data. Our analysis demonstrates that mean pairwise velocities on small scales provide a reliable and complementary cosmological probe capable of constraining parameters including neutrino mass and dark energy properties, with promising applications for upcoming peculiar velocity surveys.

This paper is organized as follows: in Sec. \ref{sec:theory}, we review the pair-conservation equation and describe the models for the nonlinear matter power spectrum included in our analysis. Section \ref{sec:methods} details the simulation datasets used and outlines the methodology and modeling framework implemented for \vot. In Sec. \ref{sec:results}, we present our results for the mean pairwise velocities in different cosmologies, along with the outcomes of the MCMC analysis used to extract cosmological constraints and assess their robustness for the \lcdm\ model. Finally, we summarize our main findings and discuss their implications in Sec. \ref{sec:conclusions}.

\section{\label{sec:theory}Pairwise velocities of large-scale structures}
This section outlines the frameworks used to model the dynamics of pairwise motions and the nonlinear clustering as an essential part of it. We first present the foundations of the model before describing the numerical approaches we employed.

\subsection{\label{subsec:modelv12}Model for pairwise velocities}
In the context of a self-gravitating system such as the large-scale structure of the Universe, the mean pairwise velocity, denoted as \vot$(r)$, is defined as the mean of the relative velocity between pairs of objects or cosmic tracers, such as galaxies or halos, separated by a distance $r$, projected along the line connecting them. It is given by
\begin{equation}\label{eq:def_v12}
v_{12}(r) = \langle[\mathbf{v}_{1}(\mathbf{r}_1) - \mathbf{v}_{2}(\mathbf{r}_2)] \cdot \hat{r}\rangle,
\end{equation}
where $\mathbf{v}_{i}$ is the peculiar velocity of the $i$-th object, $\mathbf{r}=\mathbf{r}_1-\mathbf{r}_2$ the radial vector between pairs with modulus $r=|\mathbf{r}|$ in a comoving frame. The average is computed over all pairs, and \vot\ depends only on $r$ due to isotropy. This quantity encodes the coherent motions driven by gravitational clustering and cosmic expansion. On large scales, the objects are driven apart by the Hubble flow, resulting in a negative \vot\ indicative of infall motions. 

The pair conservation equation emerges naturally from the Bogoliubov–Born–Green–Kirkwood–Yvon (BBGKY) hierarchy of equations for a self-gravitating, collisionless system of particles, where it represents the first moment of this hierarchy written as \cite{1976Ap&SS..45....3P,1983ApJ...267..465D}
\begin{equation}
    \label{eq:v12full}
    \frac{a}{3[1+\xi(x,a)]}\frac{\partial}{\partial a}\bar{\xi}(x,a) = - \frac{v_{12}(x,a)}{H(a)r},
\end{equation}
where $\xi$ is the two-point correlation function (2PCF) of the matter density field as a function of the comoving distance $x$ and the scale factor, $a$. Similarly, $\bar{\xi}$ is the volume-averaged 2PCF within a sphere of radius $x$; $r=ax$ is the proper (physical) separation and $H(a)$ is the Hubble parameter. Early attempts to solve this equation included closed-form expressions derived from perturbative expansions of the 2PCF as described in Ref. \cite{Juszkiewicz_1999}. Other approximations focused instead on the linear regime, typically for separations $x \gg 30$ \Mpch\ (see, for example Refs. \cite{1976Ap&SS..45....3P, 1980lssu.book, 1983ApJ...267..465D}). The latter approach is useful for describing the linear regime by leveraging the linear growth factor, $D(a)$, which is expressed relative to its present-day value and characterizes the growth of small density perturbations. However, this approximation breaks down in the nonlinear regime of structure formation, where perturbations become large and the simple linear scaling with the growth factor no longer applies (see Ref. \cite{2023A&ARv..31....2H} for a recent review). Some of these approximations were compared in modified gravity cosmologies for both the linear and nonlinear regimes in Ref. \cite{2024PhRvD.109l3528J}, including results from the Convolution Lagrangian Perturbation Theory  \cite{2013MNRAS.429.1674C}. Equation \eqref{eq:v12full} highlights that accurately modeling pairwise velocities in the fully nonlinear regime of structure formation requires a precise understanding of nonlinear clustering, as it explicitly depends on the time evolution of the clustering signal, which can be described either by the 2PCF or its Fourier counterpart, the matter power spectrum.

Modeling nonlinear clustering at small scales remains one of the most persistent challenges in modern cosmology. This complexity arises because accurate predictions require incorporating detailed astrophysical processes and their influence on galaxy distribution, as extensively documented in the literature. Perturbation theory methods, including modern Effective Field Theory approaches for Large Scale Structure, offer valuable analytical frameworks; however, their precision is limited in strongly nonlinear regimes \cite{2014JCAP...07..057C, 2015PhRvD..92l3007B,2022arXiv221208488I}. These approaches typically necessitate scale cutoffs and introduce multiple free parameters to account for unresolved physical effects. Alongside analytical approaches, simulation-based methods provide comprehensive insights into small-scale clustering. Recently, the advent of machine learning algorithms has enabled the development of emulators, which deliver fast and accurate predictions of statistical properties derived from simulations or theoretical models, significantly reducing computational costs. Our approach relies on the consistent numerical solution of the dynamical equations to achieve a highly accurate characterization of pairwise velocities in the nonlinear regime, with the ultimate aim of using them as probe to extract cosmological information.

\subsection{\label{subsec:modelv12Pk}Nonlinear clustering approaches}

The nonlinear modeling of the mean pairwise velocities, fundamentally hinges on the accurate description of the 2PCF over a wide range of redshifts, $z$, and cosmological parameters. This necessity arises from the explicit dependence of the expression for \vot, \refeq{v12full}, on the time evolution of the 2PCF. To ensure broad applicability, the model must remain valid across the multidimensional cosmological parameter space covered by our simulation suite, detailed in the next section. Furthermore, it must also capture the complex clustering physics inherent to small, nonlinear scales. These stringent demands, broad redshift coverage, cosmology parameter space, and accuracy in nonlinear scales substantially restrict the viable approaches for computing the nonlinear matter power spectrum. We evaluate four leading options: the semi-analytical prescriptions \code{Halofit} and HMcode\footnote{Both \code{Halofit} and \code{HMCode} were used as implemented in the \code{CLASS} code \cite{2011JCAP...07..034B}.}; alongside two high-fidelity, machine-learning-based emulators noted for their comprehensive parameter coverage and recent advent: the \code{CSST} Emulator and the \code{CosmicEmulator}. In our methodology, each approach is used to compute the nonlinear matter power spectrum, $P_{\text{NL}}(k,\, z)$, which is then Fourier transformed to obtain the 2PCF, $\xi(r,\, z)$, and then volume-averaged to produce $\bar{\xi}(r,\, z)$ for use in \refeq{v12full}.

The first approach we use to evaluate $P_{\text{NL}}$ is the \code{Halofit} prescription \cite{2003MNRAS.341.1311S, 2012ApJ...761..152T} on top of the standard linear matter power spectrum implemented within the cosmological Boltzmann solver \code{CLASS} \cite{2011JCAP...07..034B}. \code{Halofit} is an calibrated fitting function based on the halo model description, designed to approximate the nonlinear matter power spectrum. We use the updated version provided in \code{CLASS}, which builds upon earlier refinements to improve accuracy. This formulation has been tested to yield predictions within approximately 5\% for scales $k \leq 10\,h\,\mathrm{Mpc}^{-1}$ and redshifts $z \leq 2$, across $\Lambda$CDM and dynamical dark energy cosmologies \cite{2009JCAP...03..014C}. Further refinements have extended \code{Halofit}'s capabilities to account for the effects of massive neutrinos, addressing the characteristic suppression of small-scale power \cite{2012MNRAS.420.2551B}. This updated implementation sustains high fidelity in low-redshift and nonlinear regimes. Its validation against N-body simulations show discrepancies remaining below 2\% for scales of $k \sim 0.3$ $h/$Mpc at $z=1$, even with a total neutrino mass of $\sum m_{\nu} \sim 0.16$ eV \cite{2022JCAP...11..041P}.

A second semi-analytical method we employed is \code{HMCode} \cite{2015MNRAS.454.1958M,2021MNRAS.502.1401M}, which enhances the standard halo model with empirically calibrated parameters derived from high-resolution N-body simulations. This hybrid approach introduces calibrated parameters to account for key physical effects, including baryonic feedback, resulting in a more physically motivated and accurate description of clustering than purely analytical models. Both \code{Halofit} and \code{HMCode}, along with their improved variants, represent flexible and computationally efficient options, although their accuracy diminishes towards the highly nonlinear regime.

Given the known limitations of analytic approximations in the strongly nonlinear regime, we turn to simulation-based emulators. However, not all available emulators are suitable for this study. Many are limited in their range of applicability, whether in $k$, $z$, or the extent of cosmological parameter coverage. Moreover, several do not include key parameters of interest such as neutrino mass or the dark energy equation of state, which are essential for comprehensive cosmological analyses.

We chose two recent, high-fidelity emulators recognized for their broad cosmological coverage and accuracy: the \code{CosmicEmulator} \cite{2017ApJ...847...50L,2023MNRAS.520.3443M}, built from the Mira-Titan Universe simulation suite \cite{2016ApJ...820..108H}, which spans a wide range of \lcdm, $w$CDM, and $w_0w_a$CDM cosmologies, including models with massive neutrinos; and the \code{CSST Emulator} \cite{2025SCPMA..6889512C}, developed for the China Space Station Telescope (CSST) optical survey \cite{2025arXiv250704618C}. The \code{CSST Emulator}\footnote{\url{https://csst-emulator.readthedocs.io/}} is optimized for the parameter space of next-generation galaxy surveys and achieves percent-level accuracy up to $k = 10\,h\,\mathrm{Mpc}^{-1}$ across a wide range of cosmologies, including those encompassed by the \code{CosmicEmulator}. Both emulators provide fast, percent-level accurate predictions deep into the nonlinear regime by directly learning the mapping from cosmological parameters to the nonlinear matter power spectrum.

We evaluate the impact of cosmological parameters on the mean pairwise velocity across different cosmological models, including \lcdm\ and $w$CDM scenarios with massive neutrinos. The model predictions for pairwise velocities, derived from the nonlinear power spectrum methods outlined above, are rigorously compared against the \vot\ statistics directly estimated from multiple N-body simulation suites that span a broad range of cosmological parameters. The simulation data used in this analysis are described in detail in the following section.

\section{\label{sec:methods}Simulations and modeling approach}
Our analysis is based on a suite of large-scale cosmological N-body simulations. To validate the pairwise velocity model, we utilize dark-matter-only simulations from four major projects: \quijote\ \cite{2020ApJS..250....2V}, Universe N-body simulations for the Investigation of Theoretical models from galaxy surveys (hereafter \unitsim) \cite{2019MNRAS.487...48C}, Extended LEnsing PHysics using ANalytic ray Tracing runs (hereafter \elephant) \cite{2018MNRAS.476.3195C}, and \textsc{MultiDark Run1} simulations (hereafter \mdr) \cite{2012MNRAS.423.3018P}. These high-resolution datasets offer a diverse and robust framework for investigating the impact of cosmological parameters on the nonlinear clustering of matter at large scales. Table~\ref{tab:sim_params} summarizes the main features of each simulation dataset. The simulations employed are briefly described below.\\
\begin{table*}[ht]
\centering
\caption{\small{Cosmological parameters and mass resolution used in different simulation suites. In the \quijote\ simulations, bold values highlight parameters that differ from the fiducial model, with both the fiducial and varied parameter cases included. The dark matter particle mass $M_p$ is given in units of [$10^{10}$ \Msunh], and neutrino masses \Mnu\ are in units of [eV].}}
\label{tab:sim_params}
\renewcommand{\arraystretch}{1.2}
\begin{tabular}{lccccccccc}
\hline\hline
\textbf{Name} & \Om & \Ob & $h$ & $n_s$ & $\sigma_8$ & \Mnu & $w$ & $N_p$ & $M_p$ \\
\hline
\textbf{\quijote\ (fiducial)} & 0.3175 & 0.049 & 0.6711 & 0.9624 & 0.834 & 0.0 & -1.0 & 1024 & 8.206 \\
$\Omega_m^-$ & \textbf{0.3075} & 0.049 & 0.6711 & 0.9624 & 0.834 & 0.0 & -1.0 & 512 & 63.58 \\
$\Omega_m^+$ & \textbf{0.3275} & 0.049 & 0.6711 & 0.9624 & 0.834 & 0.0 & -1.0 & 512 & 67.71 \\
$h^-$ & 0.3175 & 0.049 & \textbf{0.6511} & 0.9624 & 0.834 & 0.0 & -1.0 & 512 & 65.64 \\
$h^+$ & 0.3175 & 0.049 & \textbf{0.6911} & 0.9624 & 0.834 & 0.0 & -1.0 & 512 & 65.64 \\
$\sigma_8^-$ & 0.3175 & 0.049 & 0.6711 & 0.9624 & \textbf{0.819} & 0.0 & -1.0 & 512 & 65.64 \\
$\sigma_8^+$ & 0.3175 & 0.049 & 0.6711 & 0.9624 & \textbf{0.849} & 0.0 & -1.0 & 512 & 65.64 \\
$M_\nu^+$ & 0.3175 & 0.049 & 0.6711 & 0.9624 & 0.834 & \textbf{0.2} & -1.0 & 512 & 65.64 \\
$M_\nu^{++}$ & 0.3175 & 0.049 & 0.6711 & 0.9624 & 0.834 & \textbf{0.1} & -1.0 & 512 & 65.64 \\
$w^-$ & 0.3175 & 0.049 & 0.6711 & 0.9624 & 0.834 & 0.0 & \textbf{-1.05} & 512 & 65.64 \\
$w^+$ & 0.3175 & 0.049 & 0.6711 & 0.9624 & 0.834 & 0.0 & \textbf{-0.95} & 512 & 65.64 \\
\midrule
\textbf{ELEPHANT} & 0.281 & 0.046 & 0.697 & 0.971 & 0.842 & 0.0 & -1.0 & 1024 & 7.798 \\
\midrule
\textbf{UNIT} & 0.3089 & 0.0486 & 0.6774 & 0.9667 & 0.8147 & 0.0 & -1.0 & 4096 & 0.1247 \\
\midrule
\textbf{MDR1} & 0.27 & 0.0469 & 0.70 & 0.95 & 0.82 & 0.0 & -1.0 & 2048 & 0.8721 \\
\hline\hline
\end{tabular}
\end{table*}

\textbf{ \quijote} explores a broad range of cosmological parameter variations through an extensive set of N-body simulations, including models with massive neutrinos and dynamical dark energy. The baseline simulation adopts a fiducial cosmology based on the Planck 2018 results \cite{2020A&A...641A...6P} and evolves $1024^3$ dark matter particles in a cubic box of side length 1 \Gpch, providing high-resolution insights into structure formation. In addition to the baseline, the suite systematically varies several cosmological parameters (\Om, \Ob, $h$, $n_s$, $\sigma_8$, \Mnu, and $w$), around the fiducial model as detailed in Table~\ref{tab:sim_params}. These simulations, implemented as upper and lower bounds and denoted by superscripts ``$+$'' and ``$-$'', enable a detailed assessment of how parameter variations impact both clustering and pairwise velocities. The initial conditions are generated using the second-order Lagrangian perturbation theory (2LPT) at redshift $z = 127$, with massive neutrino cosmologies treated through a particle-based approach.\\

\textbf{\unitsim} combines large simulation volumes with high resolution, offering a data set optimized for precision analyses of large-scale structure statistics within \lcdm\ cosmologies. The initial conditions are generated using the \code{FastPM} algorithm \cite{2016MNRAS.463.2273F}, based on the linear matter power spectrum for a \lcdm\ universe with cosmological parameters adopted from the Planck 2016 results \cite{2016A&A...594A..13P}. The simulations evolve $4096^3$ dark matter particles, with mass resolution of $1.2 \times 10^9\, M_\odot h^{-1}$, within a cubic volume of side length 1 \Gpch, from redshift $z=99$ to $z=0$. To minimize sample variance, the \unitsim\ simulations employ paired fixed-amplitude initial conditions.\\

\textbf{ \mdr} is a large-volume simulation designed to investigate the distribution of dark matter halos, the large-scale structure of the universe, and models of galaxy formation within a \lcdm\ scenario. The simulation adopts cosmological parameters consistent with WMAP5 results \cite{2009ApJS..180..330K} and evolves $2048^3$ particles within a cubic volume of 1 \Gpch\ per side. This corresponds to a particle mass resolution of $8.72 \times 10^9\,h^{-1}M_\odot$ and a force resolution of $7\,h^{-1}\,\mathrm{kpc}$. Initial conditions are generated at redshift $z=65$ using the Zel'dovich approximation and a linear matter power spectrum consistent with that used in the Bolshoi simulation \cite{2011ApJ...740..102K}.\\

\textbf{ \elephant} is a suite of N-body simulations designed to study the effects of modified gravity and screening mechanisms on large-scale structure, lensing, and galaxy formation. The simulations were performed using the \code{ECOSMOG} code \cite{ECOSMOG_V_1} to solve both the standard Einstein field equations and additional scalar field equations relevant to modified gravity models. Each simulation evolves $1024^3$ particles from redshift $z=49$ to $z = 0$ within a cubic volume of 1024 \Mpch, achieving a mass resolution of $7.80 \times 10^{10}\,h^{-1}M_\odot$ and a force resolution of $15\,h^{-1}\mathrm{kpc}$. The cosmological parameters are consistent with WMAP9 results \cite{2013ApJS..208...19H}. In this work, we only use the standard \lcdm\ runs to allow direct comparison with the other simulation suites.

To quantify cluster dynamics, we measure \pairvel\ directly from the complete suite of simulations described above. We use snapshots at redshifts $z = 0$, 0.5, and 1, and compute $v_{12}$ over the range $r \in [0.1, 45]\,$ \Mpch, using 50 logarithmically spaced radial bins. For each snapshot, we compute the pairwise velocities and their dispersions via direct particle-pair counting, using a random subsample containing 1\% of the dark-matter particles. This approach has been shown to achieve accuracy better than 0.5\% for separations $\gtrsim 4\,$\Mpch\ while significantly reducing the computational cost~\cite[see Appendix B1 of Ref.][]{2024MNRAS.529..360Z}. The reported $v_{12}$ values represent the mean in five independent realizations per simulation suite, with uncertainties given by the standard deviation of the mean. In the following section, we compare these simulation-based measurements with our theoretical model predictions. Specifically, we obtain $v_{12}$ for each simulation using the corresponding parameters and power spectrum prescriptions introduced earlier, and evaluate the model via the pair conservation equation after computing the necessary numerical time derivatives.

\section{\label{sec:results}Results}

\begin{figure}
\includegraphics[width=\linewidth]{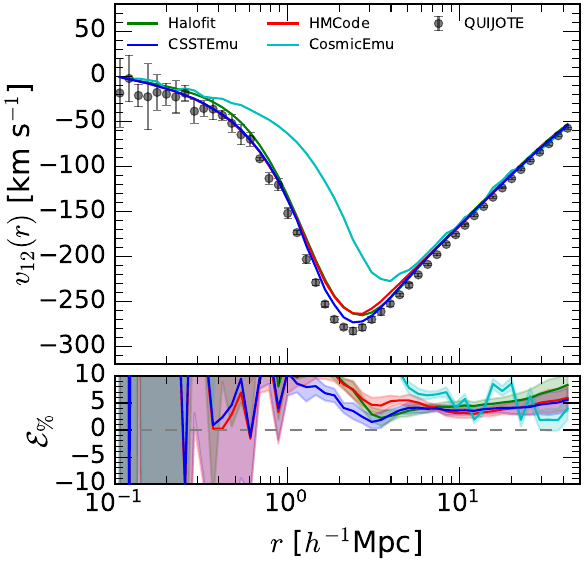}
\caption{\label{fig:v12_Pk_methods} \small{Modeling of pairwise velocities for DM tracers. Data points correspond to velocities from \quijote\ simulations at $z=0.5$. Solid lines represent the full solution of \refeq{v12full}, computed using different power spectrum generators as labeled, assuming the same fiducial cosmology as in \quijote.}}
\end{figure}

The numerical solutions for~\refeq{v12full}, obtained using different power spectra approaches, are presented in Fig.~\ref{fig:v12_Pk_methods}. The pairwise velocities, \vot, are plotted in units of $\text{km s}^{-1}$ as a function of the separation distance, which spans $0.1-40$ \Mpch\ on a logarithmic axis. This scaling allows for a detailed examination of the behavior in the nonlinear regime. For comparison, we include theoretical predictions from four key approaches: the nonlinear \code{Halofit} model (green), \code{HMCode} (red), the \code{CosmicEmulator} (cyan), and the \code{CSSTEmu} emulator (blue). The latter two are machine learning-based emulators, trained on extensive datasets of non-linear matter power spectra from N-body simulations. To evaluate their accuracy, we compare these predictions against the pairwise velocities derived from the \quijote\ fiducial cosmology, which serves as a high-fidelity numerical reference. The simulation-based pairwise velocities are represented by black dots with error bars denoting the variance across five independent realizations per snapshot. The lower panel displays the relative percentage error between the theoretical predictions and the simulation results. 

The \vot\ predicted with \code{Halofit} and \code{HMCode} exhibit a modest agreement with the \quijote\ results, although noticeable discrepancies arise at minimum velocity values around $r\approx2$ \Mpch. \code{CSSTEmu} performs especially well, with errors generally contained within a smaller range compared to the other numerical approaches. This plot highlights the challenges in accurately modeling pairwise velocities across all scales, particularly in the non-linear regime, even in the absence of tracer bias. Since the model is tested directly on dark matter particles from N-body simulations, no tracer bias assumptions are required. Additionally, the modeled $v_{12}$ depends on the time evolution of the 2PCF, which introduces further propagation of errors when discrepancies arise between the model and the simulation. The deviations observed in the plot can therefore be interpreted as a form of model bias, stemming from the intrinsic limitations of current methods in capturing small-scale nonlinear dynamics, even when employing state-of-the-art techniques. Nevertheless, these results are promising and demonstrate the potential of the methodology for further refinement. 

\begin{figure}
\includegraphics[width=\linewidth]{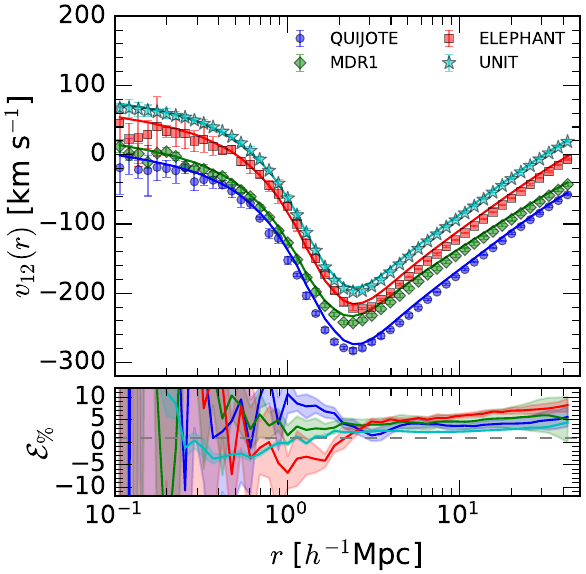}
\caption{\label{fig:v12_different_simulations} \small{Modeling of pairwise velocities for dark matter tracers. Data points correspond to simulation measurements, while solid lines show model predictions obtained with the \textsc{CSSTEmulator}. For each case, the emulator was evaluated with the corresponding cosmological parameters of the simulation and inserted into the full solution of \refeq{v12full}.}}
\end{figure}

Although we adopted the \quijote\ simulation as the primary benchmark, it is also valuable to compare the pairwise velocities behavior against other independent simulations --such as \mdr, \unitsim, and \elephant-- to assess differences in their predictions. This comparison is illustrated in Fig.~\ref{fig:v12_different_simulations}, where each simulation is represented by a distinct color and symbol, alongside the \quijote\ results. The solid lines depict the theoretical predictions from the \code{CSSTEmu}, derived from solving Eq.~\ref{eq:v12full}. Notably, all simulations exhibit a minimum in the pairwise velocities at approximately the same separation distance, $r\approx2$ \Mpch, though their absolute normalizations differ. The lower panel of the figure displays the relative percentage error between the simulations and the \code{CSSTEmu} predictions. The \quijote\ simulation shows errors below 5\% at larger scales, consistent with expectations for the quasi-linear regime. However, the error increases at smaller scales, where nonlinear effects dominate. Similar trends are observed for the other simulations, though with varying magnitudes of discrepancy, suggesting differences in their treatment of nonlinear structure formation. 

\begin{figure}
\includegraphics[width=\linewidth]{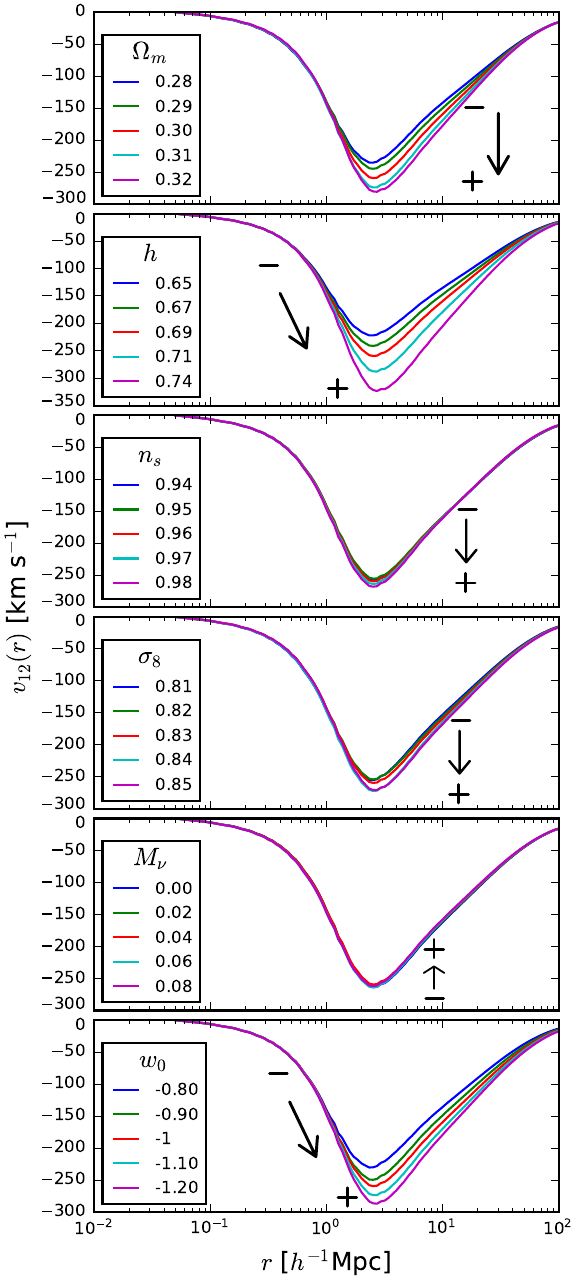}
\caption{\label{fig:v12_multicosmology} \small{Impact of cosmological parameters on the model of \vot. In each panel, one parameter from the set \{$\Omega_m$, $h$, $n_s$, $\sigma_8$, $M_\nu$, $w_0$\} is held fixed while the others are varied. Model predictions are obtained using the pair conservation alongside \code{CSSTEmu}.}}
\end{figure}

A critical aspect when modeling \vot\ is its strong dependence on cosmological parameters, particularly the matter density parameter, $\Omega_m$, and $\sigma_8$ as shown in Fig.~\ref{fig:v12_multicosmology}. Higher \Om\ enhances gravitational attraction, leading to more pronounced infall velocities (more negative \vot) at intermediate scales. This also shifts the maximum infall velocity to slightly larger scales, as the increased matter density accelerates the collapse of overdensities. Similarly, a higher $\sigma_8$ amplifies small-scale clustering, steepening the pairwise velocities profile due to stronger gravitational interactions, while the position of the maximum infall velocity remains relatively stable, as it is more sensitive to \Om. The dark energy parameter further modulate pairwise velocities dynamics, phantom equations of state $w_0<-1$ suppress late-time structure growth, reducing infall velocities primarily at large scales, where dark energy dominates. In contrast, massive neutrinos $M_\nu$ introduce a counteracting effect. As it increases, free-streaming suppresses small-scale power, weakening gravitational collapse and reducing infall velocities across all scales. 

\begin{figure}
\includegraphics[width=\linewidth]{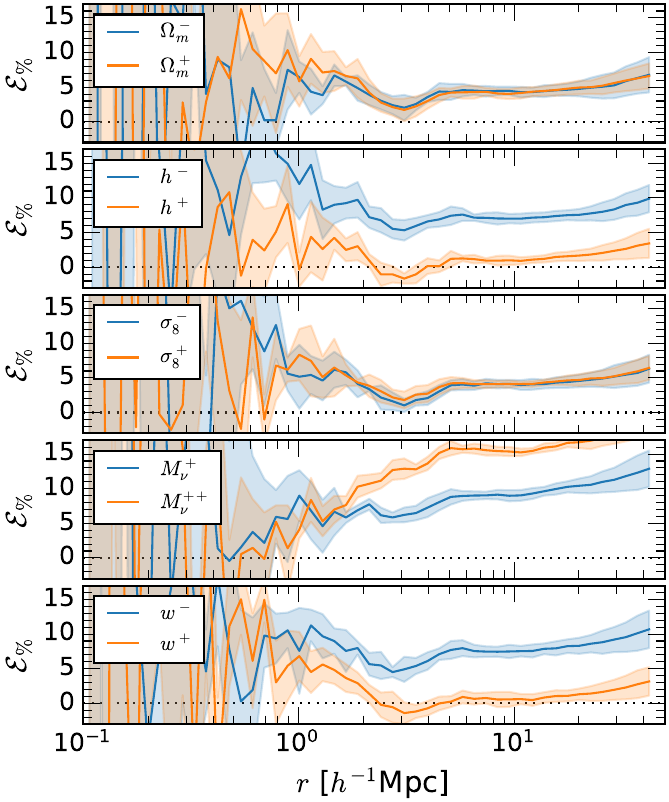}
\caption{\label{fig:v12_ratios} \small{Ratios of the pairwise velocities. Comparison of the upper and bottom limits of \quijote\ runs against the theoretical prediction.}}
\end{figure}

\begin{figure}
\includegraphics[width=\linewidth]{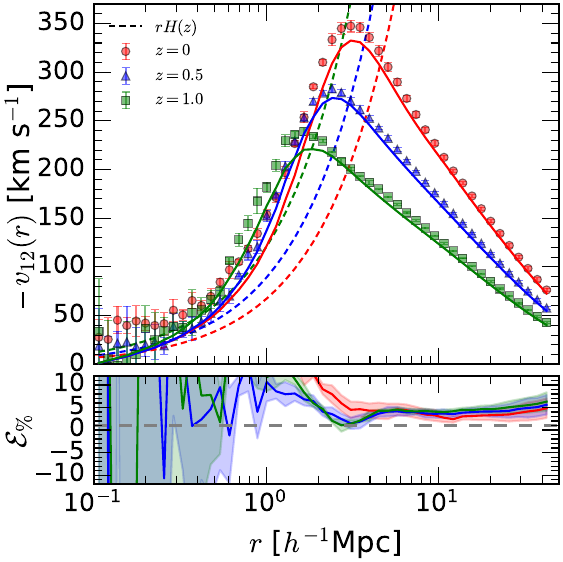}
\caption{\label{fig:stable-clustering} \small{Upper panel: Comparison of $-v_{12}$ between model predictions (solid colored lines) and simulation measurements (colored markers) against the Hubble flow, $rH(z)$ (dashed colored lines), for the fiducial \quijote\ simulation at three different redshifts, $z = 0$, $0.5$, and $1$. The intersection between the solid and dashed curves defines the stable-clustering crossing scale, $R_*$. Lower panel: relative errors between model and simulation data.}}
\end{figure}

To assess the accuracy of the pairwise velocities predictions for cosmological parameter variations, we compare the numerical outputs to the minimum and maximum values reported by the \quijote\ simulations, as shown in Fig.~\ref{fig:v12_ratios}. The results reveal that the relative percentage errors for \Om\ or $\sigma_8$, remain consistently below 10\% across the tested ranges, indicating robust agreement with the simulations. In contrast, the Hubble parameter $h$ and dark energy equation of state $w_0$ exhibit a pronounced trend, their lower bounds produce significantly larger errors in the pairwise velocities predictions. This suggests that theoretical models or emulators may struggle to accurately capture the pairwise velocities behavior for extreme values, particularly in regimes where $h$ is small or phantom-like equation-of-state values, $w_0<-1$. In addition, the uncertainty on the neutrino mass is marginally larger for a total mass of $M_\nu^{++} = 0.1$ eV than for $M_\nu^{+} = 0.2$ eV at the higher end of the separation range, indicating, as expected, the added complexity involved in constraining the properties of the lightest neutrino species. Such limitations highlight the need for further refinement in modeling the dependence of \vot\ on these parameters, especially for precision cosmology applications. Furthermore, Fig.~\ref{fig:stable-clustering} illustrates the dynamical transition at a critical separation scale across different redshifts, $z$, marking the shift from the Hubble-dominated flow to the stable-clustering regime. This transition occurs where the pairwise velocities cross the Hubble flow $rH(z)$, precisely balancing cosmic expansion against gravitational attraction. The Fig.~\ref{fig:stable-clustering} shows $-v_{12}(r,z)$ for $z=$ 0, 0.5, and 1, compared with the Hubble flow $rH$ (dashed lines). Two distinct regimes can be appreciated on the plotted scales, in the quasi-linear regime, the peculiar velocities increasingly deviate from Hubble expansion as gravitational effects become dominant, and at small scales, nonlinear interactions govern the stable clustering. The systematic shift of both the pairwise velocities minimum and critical separation to smaller scales at higher redshifts demonstrates enhanced gravitational clustering in the early universe, as expected in hierarchical structure formation. 

\begin{table}[ht]
\centering
\caption{\small{Relative errors [\%] in the predicted stable-clustering crossing scale, $R_*$, together with the associated maximum pairwise velocity, $|\hat{v}_{12}|$, and $v_{12}\!\mid_{r=R_*}$, at redshifts $z = 0$, $0.5$, and $1$ for the fiducial \quijote\ simulation.}}
\begin{tabular}{l@{\hspace{16pt}}c@{\hspace{16pt}}c@{\hspace{16pt}}c}
\hline\hline
Quantity & $z=0$ & $z=0.5$ & $z=1$ \\
\midrule
$|\hat{v}_{12}|^{\rm \quijote}$ & 347.57 & 283.16 & 238.85 \\
$|\hat{v}_{12}|^{\rm Model}$ & 332.82 & 273.72 & 221.48 \\
$\mathcal{E}(|\hat{v}_{12}|)$ & 4.43 & 3.45 & 7.84 \\
\midrule
$v_{12}\vert_{r=R_\star}^{\rm \quijote}$ & 315.83 & 271.17 & 232.85 \\
$v_{12}\vert_{r=R_\star}^{\rm Model}$ & 306.61 & 267.84 & 221.12 \\
$\mathcal{E}(|v_{12}|)$ & 3.01 & 1.24 & 5.31 \\
\midrule
$R_\star^{\rm \quijote}$ & 4.71 & 3.05 & 1.93 \\
$R_\star^{\rm Model}$ & 4.57 & 3.01 & 1.84 \\
$\mathcal{E}(|R_\star|)$ & 3.01 & 1.24 & 5.31 \\
\hline\hline
\end{tabular}
\label{tab:vmax_rstar}
\end{table}

\begin{figure}
\includegraphics[width=\linewidth]{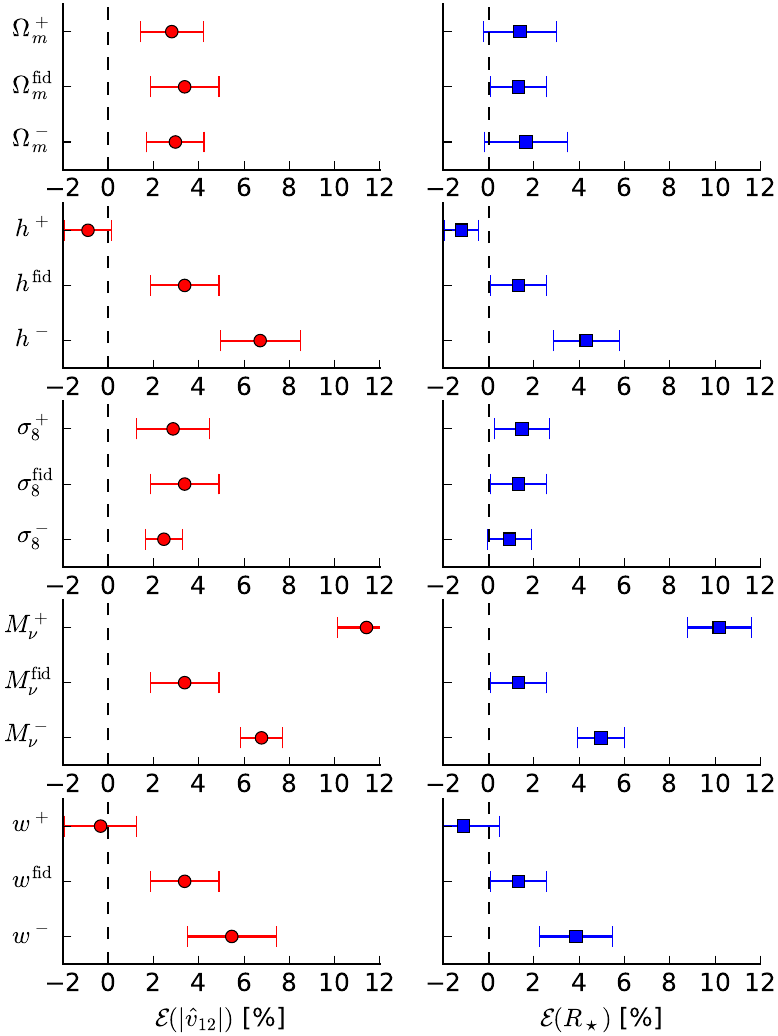}
\caption{\label{fig:v12_datamodel_comparison} \small{Relative errors in the predicted maximum infall pairwise velocity, $|\hat{v}_{12}|$, and the stable-clustering crossing scale, $R_*$, as functions of the cosmological parameters. All results are shown at $z = 0.5$.}}
\end{figure}

Table~\ref{tab:vmax_rstar} presents the errors in the maximum value of the infall pairwise velocity, $|\hat{v}_{12}|$, and the stable-clustering crossing scale, $R_*$, derived from the pair conservation model using \code{CSSTEmu} and \quijote\ simulations. The results show systematically larger errors at higher redshifts, reflecting the reduced accuracy of the emulator in earlier cosmic epochs where nonlinear structure formation becomes increasingly complex. Nonetheless, the relative error is below 10\% in all cases. Finally, Fig.~\ref{fig:v12_datamodel_comparison} quantifies the percent errors in both the maximum infall pairwise velocity and the stable-clustering crossing scale across the extreme (minimal and maximal), values of each cosmological parameter. Consistent with our earlier discussion, the errors in these quatities for \Om and $\sigma_8$ remain stable across their parameter ranges (within approximately $\pm 2\%$ across their ranges), demonstrating the robustness of these dependencies. In contrast, the remaining parameters exhibit significantly larger variations in precision between their extreme values, highlighting their more sensitive impact on the pairwise velocities and clustering transition scale. For such parameters, the error spans from $\sim8\%$ (for $w$ and $h$) to $\sim12\%$ (in $M_\nu$), emphasizing the need for precise constraints on these parameters. Thus, while \Om\ and $\sigma_8$ provide a stable foundation for modeling large-scale dynamics, the significant sensitivity of pairwise velocities and clustering scales to the remaining ones emphasizes the need to complement them with other probes and pursue a deeper theoretical understanding of these parameters in future cosmological studies.

\subsection{Inference from pairwise velocity statistics}

We implement our model within the \code{Cobaya}\footnote{\url{https://cobaya.readthedocs.io/}} framework to perform a Bayesian parameter inference using the Markov Chain Monte Carlo (MCMC) method. The core purpose of running this analysis is to efficiently sample the high-dimensional posterior distribution of cosmological parameters such $h$, \Om, and $\sigma_8$ given the likelihood defined by our data and model. We adopt the \quijote\ simulation as our fiducial cosmology. Table \ref{tab:priors} summarizes the flat priors assigned to the cosmological parameters. The subsequent MCMC analysis serves two primary purposes: first, to validate our model by comparing the inferred constraints against \quijote's true cosmology, and second, to quantify the potential of this novel probe. This process is critical for establishing robust parameter constraints, testing the impact of model assumptions, and identifying any systematic biases.

\begin{table}[t]
\centering
\renewcommand{\arraystretch}{1.15}
\caption{\small{
Flat priors assumed for the inference analysis. $\mathcal{U}[a,b]$ denotes a uniform (uninformative) prior between $a$ and $b$. The proposal width defines the standard deviation of the Gaussian proposal distribution used in the MCMC sampler.
}}
\begin{tabular}{lcc}
\hline\hline
\textbf{Parameter} & \textbf{Prior}            & \textbf{Proposal Width} \\ \hline
$h$                & $\mathcal{U}[0.62, 0.78]$ & 0.005                   \\
$\Omega_m$         & $\mathcal{U}[0.28, 0.35]$ & 0.005                   \\
$\sigma_8$         & $\mathcal{U}[0.78, 0.89]$ & 0.002                   \\ \hline\hline
\end{tabular}
\label{tab:priors}
\end{table} 

\begin{figure}
\includegraphics[width=\linewidth]{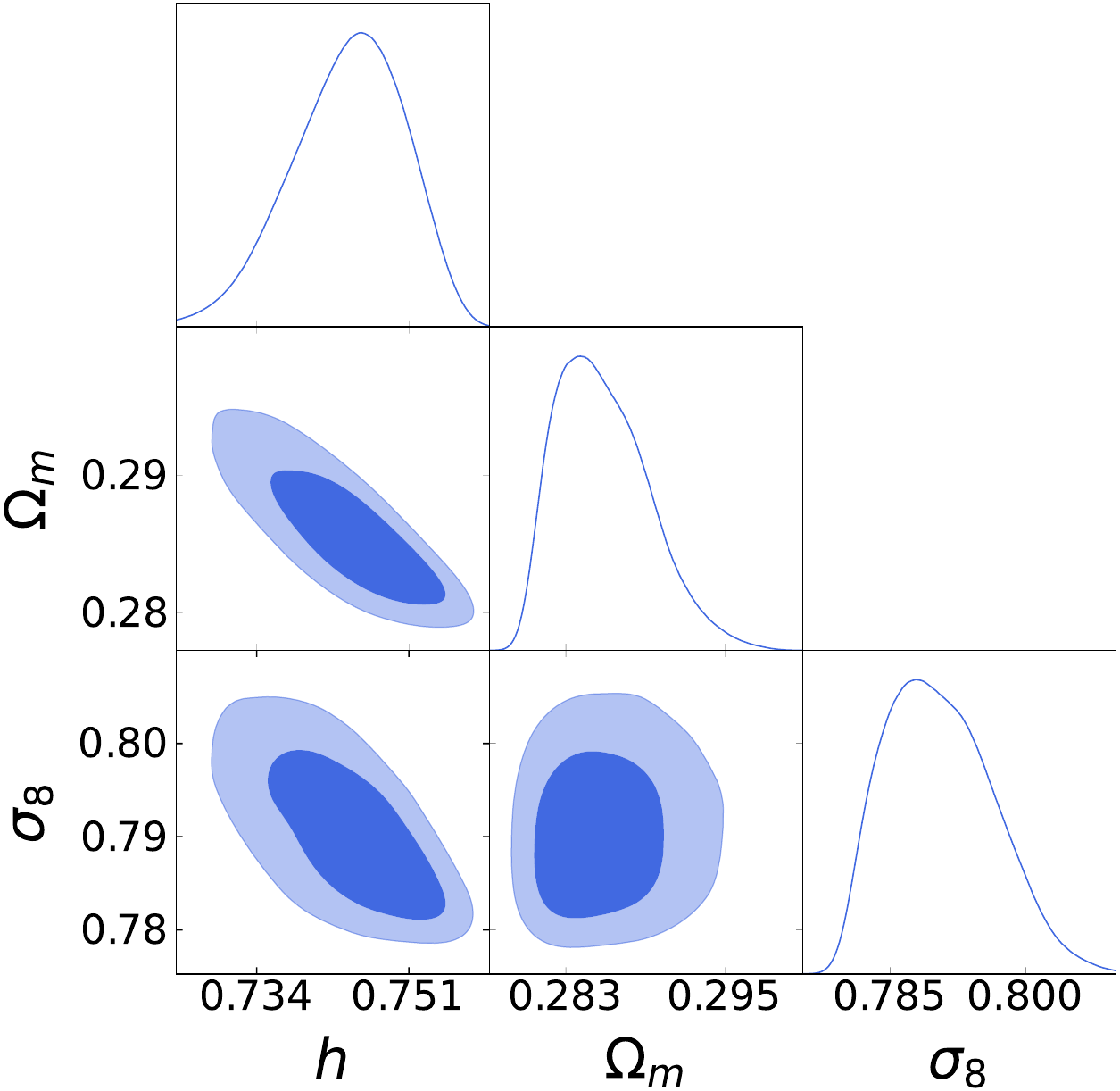}
\caption{\label{fig:v12_mcmc_triangular} \small{Confidence contours derived from the MCMC analysis of the fiducial \quijote\ data using the full model solution of \refeq{v12full}.}}
\end{figure}

\begin{figure}
\includegraphics[width=\linewidth]{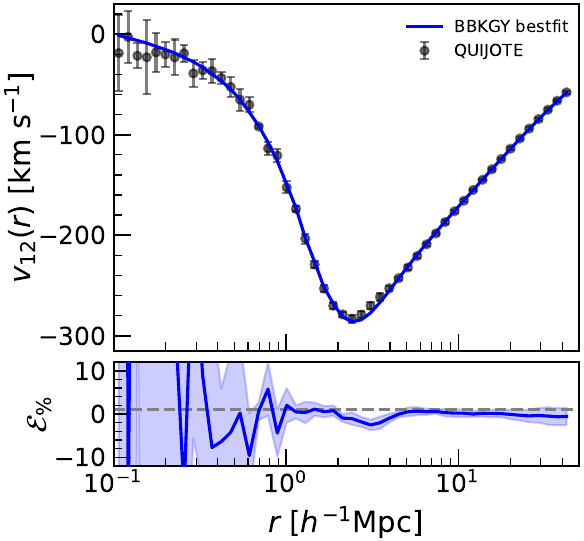}
\caption{\label{fig:v12_model_mcmc} \small{Model prediction for \vot, evaluated at the best-fit parameters from the MCMC posterior and compared with the \quijote\ simulation data.}}
\end{figure}

Figure~\ref{fig:v12_mcmc_triangular} displays the 2D posterior distributions with 1- and 2-$\sigma$ confidence regions for three cosmological parameters, while Fig.~\ref{fig:v12_model_mcmc} depicts the fit between the baseline and the pairwise velocities based on the constraints reported in Fig.~\ref{fig:v12_mcmc_triangular}. The MCMC analysis was implemented in \code{Cobaya}, yielding values of $
h = 0.7435 \pm 0.0066, \quad \Omega_{\mathrm{m}} = 0.2860 \pm 0.0035, \quad \sigma_8 = 0.7911 \pm 0.0065$.  The inferred posterior means exhibit relative deviations from the true cosmology of approximately 10.7\%, 9.9\%, and 5.1\%, respectively. These offsets would indicate the presence of unmodeled systematic biases or fundamental model limitations. We posit that these discrepancies primarily arise from the propagation of emulator inaccuracies, exacerbated by the model's nonlinear dependence on the input power spectrum.

While understanding and mitigating these systematics is crucial for future refinement, the remarkably small statistical uncertainties simultaneously underscore the probe's inherent high precision. This precision suggests immense utility in a multi-probe analysis, where it could break parameter degeneracies and significantly improve overall constraints.

Despite the current biases, the high sensitivity of pairwise velocity statistics to the underlying cosmology confirms their power as a cosmological probe, enabling the extraction of valuable information from small-scale structures. As this analysis serves as a proof of concept, future work will focus on quantifying the impact of these biases and incorporating additional physical factors to ensure robust error estimation.

\section{\label{sec:conclusions}Conclusions}
In this work, we have presented a framework to accurately model pairwise velocities across mildly non-linear and fully non-linear regimes. We demonstrated that, when clustering is properly incorporated, its time evolution yields reliable predictions for the pairwise velocities over the scales encompassed by the clustering model. We based our analysis on the equation derived in the BBGKY hierarchy and non-linear models for the power spectrum, which, via a Fourier transform, gave us the nonlinear 2PCF used as input in our core equation. This methodology stands out in that it does not require simplifications or approximations to provide a good model for \pairvel{}. Even when standard approaches to the nonlinear matter power spectrum are used as input for \refeq{v12full}, the model yields reliable predictions for \pairvel{} down to scales of a few \Mpch.

Furthermore, we have shown the sensitivity of cosmic pairwise velocities to cosmological parameters, such as \Om, $\sigma_8$, $h$, $M_\nu$, and $w$, demonstrating the potential of this approach for constraining the dark energy equation of state as well as the total neutrino mass. We showed that properly accounting for the non-linear clustering in different cosmologies can be directly input in the pair conservation model and provide a prediction for the infall velocities of pairs. In particular, the MCMC analysis performed, based the pair conservation equation and \code{CSSTEmu}, revealed discrepancies of 10.7\% in $h$, 9.9\% in \Om, and 5.1\% in $\sigma_8$, relative to the true values of the fiducial \quijote\ simulation. Future applications of this framework are extensive and must account for the complexities of the real, observed galaxy distribution. A critical next step is adding tracer bias, moving from N-body dark matter particles to the actual tracers of large-scale structure observed by spectroscopic surveys. This will involve understanding how the pairwise velocity statistics differ for bias tracers, capturing non-local cosmic evolution relative to the underlying dark matter field. Furthermore, the impact of baryonic physics, such as feedback from AGN and supernovae that can alter the distribution of matter within and around galaxies, must be integrated to achieve the precision required for next-generation cosmological data.

\begin{acknowledgments}
The authors would like to thank Baojiu Li for kindly providing the ELEPHANT simulations on behalf of \cite{2018MNRAS.476.3195C}. We are grateful with Chia-Hsun Chuang, Gustavo Y\'epez and F.-S. Kitaura for granting access to the \unitsim\ simulations.
\end{acknowledgments}

\bibliography{bibliography.bib}

\begin{thebibliography}{67}%
\makeatletter
\providecommand \@ifxundefined [1]{%
 \@ifx{#1\undefined}
}%
\providecommand \@ifnum [1]{%
 \ifnum #1\expandafter \@firstoftwo
 \else \expandafter \@secondoftwo
 \fi
}%
\providecommand \@ifx [1]{%
 \ifx #1\expandafter \@firstoftwo
 \else \expandafter \@secondoftwo
 \fi
}%
\providecommand \natexlab [1]{#1}%
\providecommand \enquote  [1]{``#1''}%
\providecommand \bibnamefont  [1]{#1}%
\providecommand \bibfnamefont [1]{#1}%
\providecommand \citenamefont [1]{#1}%
\providecommand \href@noop [0]{\@secondoftwo}%
\providecommand \href [0]{\begingroup \@sanitize@url \@href}%
\providecommand \@href[1]{\@@startlink{#1}\@@href}%
\providecommand \@@href[1]{\endgroup#1\@@endlink}%
\providecommand \@sanitize@url [0]{\catcode `\\12\catcode `\$12\catcode
  `\&12\catcode `\#12\catcode `\^12\catcode `\_12\catcode `\%12\relax}%
\providecommand \@@startlink[1]{}%
\providecommand \@@endlink[0]{}%
\providecommand \url  [0]{\begingroup\@sanitize@url \@url }%
\providecommand \@url [1]{\endgroup\@href {#1}{\urlprefix }}%
\providecommand \urlprefix  [0]{URL }%
\providecommand \Eprint [0]{\href }%
\providecommand \doibase [0]{http://dx.doi.org/}%
\providecommand \selectlanguage [0]{\@gobble}%
\providecommand \bibinfo  [0]{\@secondoftwo}%
\providecommand \bibfield  [0]{\@secondoftwo}%
\providecommand \translation [1]{[#1]}%
\providecommand \BibitemOpen [0]{}%
\providecommand \bibitemStop [0]{}%
\providecommand \bibitemNoStop [0]{.\EOS\space}%
\providecommand \EOS [0]{\spacefactor3000\relax}%
\providecommand \BibitemShut  [1]{\csname bibitem#1\endcsname}%
\let\auto@bib@innerbib\@empty
\bibitem [{\citenamefont {{Andersen}}\ \emph {et~al.}(2016)\citenamefont
  {{Andersen}}, \citenamefont {{Davis}},\ and\ \citenamefont
  {{Howlett}}}]{2016MNRAS.463.4083A}%
  \BibitemOpen
  \bibfield  {author} {\bibinfo {author} {\bibfnamefont {P.}~\bibnamefont
  {{Andersen}}}, \bibinfo {author} {\bibfnamefont {T.~M.}\ \bibnamefont
  {{Davis}}}, \ and\ \bibinfo {author} {\bibfnamefont {C.}~\bibnamefont
  {{Howlett}}},\ }\bibfield  {title} {\enquote {\bibinfo {title} {{Cosmology
  with peculiar velocities: observational effects}},}\ }\href {\doibase
  10.1093/mnras/stw2252} {\bibfield  {journal} {\bibinfo  {journal} {\mnras}\
  }\textbf {\bibinfo {volume} {463}},\ \bibinfo {pages} {4083} (\bibinfo {year}
  {2016})},\ \Eprint {http://arxiv.org/abs/1609.04022}{arXiv:1609.04022
  [astro-ph.CO]}\BibitemShut {NoStop}%
\bibitem [{\citenamefont {{Hudson}}\ and\ \citenamefont
  {{Turnbull}}(2012)}]{2012ApJ...751L..30H}%
  \BibitemOpen
  \bibfield  {author} {\bibinfo {author} {\bibfnamefont {M.~J.}\ \bibnamefont
  {{Hudson}}}\ and\ \bibinfo {author} {\bibfnamefont {S.~J.}\ \bibnamefont
  {{Turnbull}}},\ }\bibfield  {title} {\enquote {\bibinfo {title} {{The Growth
  Rate of Cosmic Structure from Peculiar Velocities at Low and High
  Redshifts}},}\ }\href {\doibase 10.1088/2041-8205/751/2/L30} {\bibfield
  {journal} {\bibinfo  {journal} {\apjl}\ }\textbf {\bibinfo {volume} {751}},\
  \bibinfo {eid} {L30} (\bibinfo {year} {2012})},\ \Eprint
  {http://arxiv.org/abs/1203.4814}{arXiv:1203.4814 [astro-ph.CO]}\BibitemShut
  {NoStop}%
\bibitem [{\citenamefont {{Koda}}\ \emph {et~al.}(2014)\citenamefont {{Koda}},
  \citenamefont {{Blake}}, \citenamefont {{Davis}}, \citenamefont {{Magoulas}},
  \citenamefont {{Springob}}, \citenamefont {{Scrimgeour}}, \citenamefont
  {{Johnson}}, \citenamefont {{Poole}},\ and\ \citenamefont
  {{Staveley-Smith}}}]{2014MNRAS.445.4267K}%
  \BibitemOpen
  \bibfield  {author} {\bibinfo {author} {\bibfnamefont {J.}~\bibnamefont
  {{Koda}}}, \bibinfo {author} {\bibfnamefont {C.}~\bibnamefont {{Blake}}},
  \bibinfo {author} {\bibfnamefont {T.}~\bibnamefont {{Davis}}}, \bibinfo
  {author} {\bibfnamefont {C.}~\bibnamefont {{Magoulas}}}, \bibinfo {author}
  {\bibfnamefont {C.~M.}\ \bibnamefont {{Springob}}}, \bibinfo {author}
  {\bibfnamefont {M.}~\bibnamefont {{Scrimgeour}}}, \bibinfo {author}
  {\bibfnamefont {A.}~\bibnamefont {{Johnson}}}, \bibinfo {author}
  {\bibfnamefont {G.~B.}\ \bibnamefont {{Poole}}}, \ and\ \bibinfo {author}
  {\bibfnamefont {L.}~\bibnamefont {{Staveley-Smith}}},\ }\bibfield  {title}
  {\enquote {\bibinfo {title} {{Are peculiar velocity surveys competitive as a
  cosmological probe?}}}\ }\href {\doibase 10.1093/mnras/stu1610} {\bibfield
  {journal} {\bibinfo  {journal} {\mnras}\ }\textbf {\bibinfo {volume} {445}},\
  \bibinfo {pages} {4267} (\bibinfo {year} {2014})},\ \Eprint
  {http://arxiv.org/abs/1312.1022}{arXiv:1312.1022 [astro-ph.CO]}\BibitemShut
  {NoStop}%
\bibitem [{\citenamefont {{Weinberg}}\ \emph {et~al.}(2013)\citenamefont
  {{Weinberg}}, \citenamefont {{Mortonson}}, \citenamefont {{Eisenstein}},
  \citenamefont {{Hirata}}, \citenamefont {{Riess}},\ and\ \citenamefont
  {{Rozo}}}]{2013PhR...530...87W}%
  \BibitemOpen
  \bibfield  {author} {\bibinfo {author} {\bibfnamefont {D.~H.}\ \bibnamefont
  {{Weinberg}}}, \bibinfo {author} {\bibfnamefont {M.~J.}\ \bibnamefont
  {{Mortonson}}}, \bibinfo {author} {\bibfnamefont {D.~J.}\ \bibnamefont
  {{Eisenstein}}}, \bibinfo {author} {\bibfnamefont {C.}~\bibnamefont
  {{Hirata}}}, \bibinfo {author} {\bibfnamefont {A.~G.}\ \bibnamefont
  {{Riess}}}, \ and\ \bibinfo {author} {\bibfnamefont {E.}~\bibnamefont
  {{Rozo}}},\ }\bibfield  {title} {\enquote {\bibinfo {title} {{Observational
  probes of cosmic acceleration}},}\ }\href {\doibase
  10.1016/j.physrep.2013.05.001} {\bibfield  {journal} {\bibinfo  {journal}
  {\physrep}\ }\textbf {\bibinfo {volume} {530}},\ \bibinfo {pages} {87}
  (\bibinfo {year} {2013})},\ \Eprint
  {http://arxiv.org/abs/1201.2434}{arXiv:1201.2434 [astro-ph.CO]}\BibitemShut
  {NoStop}%
\bibitem [{\citenamefont {{Hildebrandt}}\ \emph {et~al.}(2017)\citenamefont
  {{Hildebrandt}}, \citenamefont {{Viola}}, \citenamefont {{Heymans}},
  \citenamefont {{Joudaki}}, \citenamefont {{Kuijken}}, \citenamefont
  {{Blake}}, \citenamefont {{Erben}}, \citenamefont {{Joachimi}}, \citenamefont
  {{Klaes}}, \citenamefont {{Miller}} \emph {et~al.}}]{2017MNRAS.465.1454H}%
  \BibitemOpen
  \bibfield  {author} {\bibinfo {author} {\bibfnamefont {H.}~\bibnamefont
  {{Hildebrandt}}}, \bibinfo {author} {\bibfnamefont {M.}~\bibnamefont
  {{Viola}}}, \bibinfo {author} {\bibfnamefont {C.}~\bibnamefont {{Heymans}}},
  \bibinfo {author} {\bibfnamefont {S.}~\bibnamefont {{Joudaki}}}, \bibinfo
  {author} {\bibfnamefont {K.}~\bibnamefont {{Kuijken}}}, \bibinfo {author}
  {\bibfnamefont {C.}~\bibnamefont {{Blake}}}, \bibinfo {author} {\bibfnamefont
  {T.}~\bibnamefont {{Erben}}}, \bibinfo {author} {\bibfnamefont
  {B.}~\bibnamefont {{Joachimi}}}, \bibinfo {author} {\bibfnamefont
  {D.}~\bibnamefont {{Klaes}}}, \bibinfo {author} {\bibfnamefont
  {L.}~\bibnamefont {{Miller}}},  \emph {et~al.},\ }\bibfield  {title}
  {\enquote {\bibinfo {title} {{KiDS-450: cosmological parameter constraints
  from tomographic weak gravitational lensing}},}\ }\href {\doibase
  10.1093/mnras/stw2805} {\bibfield  {journal} {\bibinfo  {journal} {\mnras}\
  }\textbf {\bibinfo {volume} {465}},\ \bibinfo {pages} {1454} (\bibinfo {year}
  {2017})},\ \Eprint {http://arxiv.org/abs/1606.05338}{arXiv:1606.05338
  [astro-ph.CO]}\BibitemShut {NoStop}%
\bibitem [{\citenamefont {{Planck Collaboration}}\ \emph
  {et~al.}(2020)\citenamefont {{Planck Collaboration}}, \citenamefont
  {{Aghanim}}, \citenamefont {{Akrami}}, \citenamefont {{Ashdown}},
  \citenamefont {{Aumont}}, \citenamefont {{Baccigalupi}}, \citenamefont
  {{Ballardini}}, \citenamefont {{Banday}}, \citenamefont {{Barreiro}},
  \citenamefont {{Bartolo}} \emph {et~al.}}]{2020A&A...641A...6P}%
  \BibitemOpen
  \bibfield  {author} {\bibinfo {author} {\bibnamefont {{Planck
  Collaboration}}}, \bibinfo {author} {\bibfnamefont {N.}~\bibnamefont
  {{Aghanim}}}, \bibinfo {author} {\bibfnamefont {Y.}~\bibnamefont {{Akrami}}},
  \bibinfo {author} {\bibfnamefont {M.}~\bibnamefont {{Ashdown}}}, \bibinfo
  {author} {\bibfnamefont {J.}~\bibnamefont {{Aumont}}}, \bibinfo {author}
  {\bibfnamefont {C.}~\bibnamefont {{Baccigalupi}}}, \bibinfo {author}
  {\bibfnamefont {M.}~\bibnamefont {{Ballardini}}}, \bibinfo {author}
  {\bibfnamefont {A.~J.}\ \bibnamefont {{Banday}}}, \bibinfo {author}
  {\bibfnamefont {R.~B.}\ \bibnamefont {{Barreiro}}}, \bibinfo {author}
  {\bibfnamefont {N.}~\bibnamefont {{Bartolo}}},  \emph {et~al.},\ }\bibfield
  {title} {\enquote {\bibinfo {title} {{Planck 2018 results. VI. Cosmological
  parameters}},}\ }\href {\doibase 10.1051/0004-6361/201833910} {\bibfield
  {journal} {\bibinfo  {journal} {\aap}\ }\textbf {\bibinfo {volume} {641}},\
  \bibinfo {eid} {A6} (\bibinfo {year} {2020})},\ \Eprint
  {http://arxiv.org/abs/1807.06209}{arXiv:1807.06209 [astro-ph.CO]}\BibitemShut
  {NoStop}%
\bibitem [{\citenamefont {{Riess}}\ \emph {et~al.}(2022)\citenamefont
  {{Riess}}, \citenamefont {{Yuan}}, \citenamefont {{Macri}}, \citenamefont
  {{Scolnic}}, \citenamefont {{Brout}}, \citenamefont {{Casertano}},
  \citenamefont {{Jones}}, \citenamefont {{Murakami}}, \citenamefont {{Anand}},
  \citenamefont {{Breuval}} \emph {et~al.}}]{2022ApJ...934L...7R}%
  \BibitemOpen
  \bibfield  {author} {\bibinfo {author} {\bibfnamefont {A.~G.}\ \bibnamefont
  {{Riess}}}, \bibinfo {author} {\bibfnamefont {W.}~\bibnamefont {{Yuan}}},
  \bibinfo {author} {\bibfnamefont {L.~M.}\ \bibnamefont {{Macri}}}, \bibinfo
  {author} {\bibfnamefont {D.}~\bibnamefont {{Scolnic}}}, \bibinfo {author}
  {\bibfnamefont {D.}~\bibnamefont {{Brout}}}, \bibinfo {author} {\bibfnamefont
  {S.}~\bibnamefont {{Casertano}}}, \bibinfo {author} {\bibfnamefont {D.~O.}\
  \bibnamefont {{Jones}}}, \bibinfo {author} {\bibfnamefont {Y.}~\bibnamefont
  {{Murakami}}}, \bibinfo {author} {\bibfnamefont {G.~S.}\ \bibnamefont
  {{Anand}}}, \bibinfo {author} {\bibfnamefont {L.}~\bibnamefont {{Breuval}}},
  \emph {et~al.},\ }\bibfield  {title} {\enquote {\bibinfo {title} {{A
  Comprehensive Measurement of the Local Value of the Hubble Constant with 1 km
  s$^{-1}$ Mpc$^{-1}$ Uncertainty from the Hubble Space Telescope and the SH0ES
  Team}},}\ }\href {\doibase 10.3847/2041-8213/ac5c5b} {\bibfield  {journal}
  {\bibinfo  {journal} {\apjl}\ }\textbf {\bibinfo {volume} {934}},\ \bibinfo
  {eid} {L7} (\bibinfo {year} {2022})},\ \Eprint
  {http://arxiv.org/abs/2112.04510}{arXiv:2112.04510 [astro-ph.CO]}\BibitemShut
  {NoStop}%
\bibitem [{\citenamefont {{Gil-Mar{\'\i}n}}\ \emph {et~al.}(2012)\citenamefont
  {{Gil-Mar{\'\i}n}}, \citenamefont {{Wagner}}, \citenamefont {{Verde}},
  \citenamefont {{Porciani}},\ and\ \citenamefont
  {{Jimenez}}}]{2012JCAP...11..029G}%
  \BibitemOpen
  \bibfield  {author} {\bibinfo {author} {\bibfnamefont {H.}~\bibnamefont
  {{Gil-Mar{\'\i}n}}}, \bibinfo {author} {\bibfnamefont {C.}~\bibnamefont
  {{Wagner}}}, \bibinfo {author} {\bibfnamefont {L.}~\bibnamefont {{Verde}}},
  \bibinfo {author} {\bibfnamefont {C.}~\bibnamefont {{Porciani}}}, \ and\
  \bibinfo {author} {\bibfnamefont {R.}~\bibnamefont {{Jimenez}}},\ }\bibfield
  {title} {\enquote {\bibinfo {title} {{Perturbation theory approach for the
  power spectrum: from dark matter in real space to massive haloes in redshift
  space}},}\ }\href {\doibase 10.1088/1475-7516/2012/11/029} {\bibfield
  {journal} {\bibinfo  {journal} {\jcap}\ }\textbf {\bibinfo {volume} {2012}},\
  \bibinfo {eid} {029} (\bibinfo {year} {2012})},\ \Eprint
  {http://arxiv.org/abs/1209.3771}{arXiv:1209.3771 [astro-ph.CO]}\BibitemShut
  {NoStop}%
\bibitem [{\citenamefont {{Taruya}}\ \emph {et~al.}(2010)\citenamefont
  {{Taruya}}, \citenamefont {{Nishimichi}},\ and\ \citenamefont
  {{Saito}}}]{2010PhRvD..82f3522T}%
  \BibitemOpen
  \bibfield  {author} {\bibinfo {author} {\bibfnamefont {A.}~\bibnamefont
  {{Taruya}}}, \bibinfo {author} {\bibfnamefont {T.}~\bibnamefont
  {{Nishimichi}}}, \ and\ \bibinfo {author} {\bibfnamefont {S.}~\bibnamefont
  {{Saito}}},\ }\bibfield  {title} {\enquote {\bibinfo {title} {{Baryon
  acoustic oscillations in 2D: Modeling redshift-space power spectrum from
  perturbation theory}},}\ }\href {\doibase 10.1103/PhysRevD.82.063522}
  {\bibfield  {journal} {\bibinfo  {journal} {\prd}\ }\textbf {\bibinfo
  {volume} {82}},\ \bibinfo {eid} {063522} (\bibinfo {year} {2010})},\ \Eprint
  {http://arxiv.org/abs/1006.0699}{arXiv:1006.0699 [astro-ph.CO]}\BibitemShut
  {NoStop}%
\bibitem [{\citenamefont {{Scoccimarro}}(2004)}]{2004PhRvD..70h3007S}%
  \BibitemOpen
  \bibfield  {author} {\bibinfo {author} {\bibfnamefont {R.}~\bibnamefont
  {{Scoccimarro}}},\ }\bibfield  {title} {\enquote {\bibinfo {title}
  {{Redshift-space distortions, pairwise velocities, and nonlinearities}},}\
  }\href {\doibase 10.1103/PhysRevD.70.083007} {\bibfield  {journal} {\bibinfo
  {journal} {\prd}\ }\textbf {\bibinfo {volume} {70}},\ \bibinfo {eid} {083007}
  (\bibinfo {year} {2004})},\ \Eprint
  {http://arxiv.org/abs/astro-ph/0407214}{arXiv:astro-ph/0407214
  [astro-ph]}\BibitemShut {NoStop}%
\bibitem [{\citenamefont {{Bernardeau}}\ \emph {et~al.}(2002)\citenamefont
  {{Bernardeau}}, \citenamefont {{Colombi}}, \citenamefont {{Gazta{\~n}aga}},\
  and\ \citenamefont {{Scoccimarro}}}]{2002PhR...367....1B}%
  \BibitemOpen
  \bibfield  {author} {\bibinfo {author} {\bibfnamefont {F.}~\bibnamefont
  {{Bernardeau}}}, \bibinfo {author} {\bibfnamefont {S.}~\bibnamefont
  {{Colombi}}}, \bibinfo {author} {\bibfnamefont {E.}~\bibnamefont
  {{Gazta{\~n}aga}}}, \ and\ \bibinfo {author} {\bibfnamefont {R.}~\bibnamefont
  {{Scoccimarro}}},\ }\bibfield  {title} {\enquote {\bibinfo {title}
  {{Large-scale structure of the Universe and cosmological perturbation
  theory}},}\ }\href {\doibase 10.1016/S0370-1573(02)00135-7} {\bibfield
  {journal} {\bibinfo  {journal} {\physrep}\ }\textbf {\bibinfo {volume}
  {367}},\ \bibinfo {pages} {1} (\bibinfo {year} {2002})},\ \Eprint
  {http://arxiv.org/abs/astro-ph/0112551}{arXiv:astro-ph/0112551
  [astro-ph]}\BibitemShut {NoStop}%
\bibitem [{\citenamefont {{Taruya}}(2014)}]{2014ascl.soft04012T}%
  \BibitemOpen
  \bibfield  {author} {\bibinfo {author} {\bibfnamefont {A.}~\bibnamefont
  {{Taruya}}},\ }\href@noop {} {\enquote {\bibinfo {title} {{RegPT: Regularized
  cosmological power spectrum}},}\ }\bibinfo {howpublished} {Astrophysics
  Source Code Library, record ascl:1404.012} (\bibinfo {year} {2014}),\ \Eprint
  {http://arxiv.org/abs/1404.012}{ascl:1404.012}\BibitemShut {NoStop}%
\bibitem [{\citenamefont {{Baumann}}\ \emph {et~al.}(2012)\citenamefont
  {{Baumann}}, \citenamefont {{Nicolis}}, \citenamefont {{Senatore}},\ and\
  \citenamefont {{Zaldarriaga}}}]{2012JCAP...07..051B}%
  \BibitemOpen
  \bibfield  {author} {\bibinfo {author} {\bibfnamefont {D.}~\bibnamefont
  {{Baumann}}}, \bibinfo {author} {\bibfnamefont {A.}~\bibnamefont
  {{Nicolis}}}, \bibinfo {author} {\bibfnamefont {L.}~\bibnamefont
  {{Senatore}}}, \ and\ \bibinfo {author} {\bibfnamefont {M.}~\bibnamefont
  {{Zaldarriaga}}},\ }\bibfield  {title} {\enquote {\bibinfo {title}
  {{Cosmological non-linearities as an effective fluid}},}\ }\href {\doibase
  10.1088/1475-7516/2012/07/051} {\bibfield  {journal} {\bibinfo  {journal}
  {\jcap}\ }\textbf {\bibinfo {volume} {2012}},\ \bibinfo {eid} {051} (\bibinfo
  {year} {2012})},\ \Eprint {http://arxiv.org/abs/1004.2488}{arXiv:1004.2488
  [astro-ph.CO]}\BibitemShut {NoStop}%
\bibitem [{\citenamefont {{Carrasco}}\ \emph {et~al.}(2012)\citenamefont
  {{Carrasco}}, \citenamefont {{Hertzberg}},\ and\ \citenamefont
  {{Senatore}}}]{2012JHEP...09..082C}%
  \BibitemOpen
  \bibfield  {author} {\bibinfo {author} {\bibfnamefont {J.~J.~M.}\
  \bibnamefont {{Carrasco}}}, \bibinfo {author} {\bibfnamefont {M.~P.}\
  \bibnamefont {{Hertzberg}}}, \ and\ \bibinfo {author} {\bibfnamefont
  {L.}~\bibnamefont {{Senatore}}},\ }\bibfield  {title} {\enquote {\bibinfo
  {title} {{The effective field theory of cosmological large scale
  structures}},}\ }\href {\doibase 10.1007/JHEP09(2012)082} {\bibfield
  {journal} {\bibinfo  {journal} {Journal of High Energy Physics}\ }\textbf
  {\bibinfo {volume} {2012}},\ \bibinfo {eid} {82} (\bibinfo {year} {2012})},\
  \Eprint {http://arxiv.org/abs/1206.2926}{arXiv:1206.2926
  [astro-ph.CO]}\BibitemShut {NoStop}%
\bibitem [{\citenamefont {{Smith}}\ \emph {et~al.}(2003)\citenamefont
  {{Smith}}, \citenamefont {{Peacock}}, \citenamefont {{Jenkins}},
  \citenamefont {{White}}, \citenamefont {{Frenk}}, \citenamefont {{Pearce}},
  \citenamefont {{Thomas}}, \citenamefont {{Efstathiou}},\ and\ \citenamefont
  {{Couchman}}}]{2003MNRAS.341.1311S}%
  \BibitemOpen
  \bibfield  {author} {\bibinfo {author} {\bibfnamefont {R.~E.}\ \bibnamefont
  {{Smith}}}, \bibinfo {author} {\bibfnamefont {J.~A.}\ \bibnamefont
  {{Peacock}}}, \bibinfo {author} {\bibfnamefont {A.}~\bibnamefont
  {{Jenkins}}}, \bibinfo {author} {\bibfnamefont {S.~D.~M.}\ \bibnamefont
  {{White}}}, \bibinfo {author} {\bibfnamefont {C.~S.}\ \bibnamefont
  {{Frenk}}}, \bibinfo {author} {\bibfnamefont {F.~R.}\ \bibnamefont
  {{Pearce}}}, \bibinfo {author} {\bibfnamefont {P.~A.}\ \bibnamefont
  {{Thomas}}}, \bibinfo {author} {\bibfnamefont {G.}~\bibnamefont
  {{Efstathiou}}}, \ and\ \bibinfo {author} {\bibfnamefont {H.~M.~P.}\
  \bibnamefont {{Couchman}}},\ }\bibfield  {title} {\enquote {\bibinfo {title}
  {{Stable clustering, the halo model and non-linear cosmological power
  spectra}},}\ }\href {\doibase 10.1046/j.1365-8711.2003.06503.x} {\bibfield
  {journal} {\bibinfo  {journal} {\mnras}\ }\textbf {\bibinfo {volume} {341}},\
  \bibinfo {pages} {1311} (\bibinfo {year} {2003})},\ \Eprint
  {http://arxiv.org/abs/astro-ph/0207664}{arXiv:astro-ph/0207664
  [astro-ph]}\BibitemShut {NoStop}%
\bibitem [{\citenamefont {{Takahashi}}\ \emph {et~al.}(2012)\citenamefont
  {{Takahashi}}, \citenamefont {{Sato}}, \citenamefont {{Nishimichi}},
  \citenamefont {{Taruya}},\ and\ \citenamefont
  {{Oguri}}}]{2012ApJ...761..152T}%
  \BibitemOpen
  \bibfield  {author} {\bibinfo {author} {\bibfnamefont {R.}~\bibnamefont
  {{Takahashi}}}, \bibinfo {author} {\bibfnamefont {M.}~\bibnamefont {{Sato}}},
  \bibinfo {author} {\bibfnamefont {T.}~\bibnamefont {{Nishimichi}}}, \bibinfo
  {author} {\bibfnamefont {A.}~\bibnamefont {{Taruya}}}, \ and\ \bibinfo
  {author} {\bibfnamefont {M.}~\bibnamefont {{Oguri}}},\ }\bibfield  {title}
  {\enquote {\bibinfo {title} {{Revising the Halofit Model for the Nonlinear
  Matter Power Spectrum}},}\ }\href {\doibase 10.1088/0004-637X/761/2/152}
  {\bibfield  {journal} {\bibinfo  {journal} {\apj}\ }\textbf {\bibinfo
  {volume} {761}},\ \bibinfo {eid} {152} (\bibinfo {year} {2012})},\ \Eprint
  {http://arxiv.org/abs/1208.2701}{arXiv:1208.2701 [astro-ph.CO]}\BibitemShut
  {NoStop}%
\bibitem [{\citenamefont {{Mead}}\ \emph {et~al.}(2015)\citenamefont {{Mead}},
  \citenamefont {{Peacock}}, \citenamefont {{Heymans}}, \citenamefont
  {{Joudaki}},\ and\ \citenamefont {{Heavens}}}]{2015MNRAS.454.1958M}%
  \BibitemOpen
  \bibfield  {author} {\bibinfo {author} {\bibfnamefont {A.~J.}\ \bibnamefont
  {{Mead}}}, \bibinfo {author} {\bibfnamefont {J.~A.}\ \bibnamefont
  {{Peacock}}}, \bibinfo {author} {\bibfnamefont {C.}~\bibnamefont
  {{Heymans}}}, \bibinfo {author} {\bibfnamefont {S.}~\bibnamefont
  {{Joudaki}}}, \ and\ \bibinfo {author} {\bibfnamefont {A.~F.}\ \bibnamefont
  {{Heavens}}},\ }\bibfield  {title} {\enquote {\bibinfo {title} {{An accurate
  halo model for fitting non-linear cosmological power spectra and baryonic
  feedback models}},}\ }\href {\doibase 10.1093/mnras/stv2036} {\bibfield
  {journal} {\bibinfo  {journal} {\mnras}\ }\textbf {\bibinfo {volume} {454}},\
  \bibinfo {pages} {1958} (\bibinfo {year} {2015})},\ \Eprint
  {http://arxiv.org/abs/1505.07833}{arXiv:1505.07833 [astro-ph.CO]}\BibitemShut
  {NoStop}%
\bibitem [{\citenamefont {{Heitmann}}\ \emph {et~al.}(2014)\citenamefont
  {{Heitmann}}, \citenamefont {{Lawrence}}, \citenamefont {{Kwan}},
  \citenamefont {{Habib}},\ and\ \citenamefont
  {{Higdon}}}]{2014ApJ...780..111H}%
  \BibitemOpen
  \bibfield  {author} {\bibinfo {author} {\bibfnamefont {K.}~\bibnamefont
  {{Heitmann}}}, \bibinfo {author} {\bibfnamefont {E.}~\bibnamefont
  {{Lawrence}}}, \bibinfo {author} {\bibfnamefont {J.}~\bibnamefont {{Kwan}}},
  \bibinfo {author} {\bibfnamefont {S.}~\bibnamefont {{Habib}}}, \ and\
  \bibinfo {author} {\bibfnamefont {D.}~\bibnamefont {{Higdon}}},\ }\bibfield
  {title} {\enquote {\bibinfo {title} {{The Coyote Universe Extended: Precision
  Emulation of the Matter Power Spectrum}},}\ }\href {\doibase
  10.1088/0004-637X/780/1/111} {\bibfield  {journal} {\bibinfo  {journal}
  {\apj}\ }\textbf {\bibinfo {volume} {780}},\ \bibinfo {eid} {111} (\bibinfo
  {year} {2014})},\ \Eprint {http://arxiv.org/abs/1304.7849}{arXiv:1304.7849
  [astro-ph.CO]}\BibitemShut {NoStop}%
\bibitem [{\citenamefont {{Winther}}\ \emph {et~al.}(2019)\citenamefont
  {{Winther}}, \citenamefont {{Casas}}, \citenamefont {{Baldi}}, \citenamefont
  {{Koyama}}, \citenamefont {{Li}}, \citenamefont {{Lombriser}},\ and\
  \citenamefont {{Zhao}}}]{2019PhRvD.100l3540W}%
  \BibitemOpen
  \bibfield  {author} {\bibinfo {author} {\bibfnamefont {H.~A.}\ \bibnamefont
  {{Winther}}}, \bibinfo {author} {\bibfnamefont {S.}~\bibnamefont {{Casas}}},
  \bibinfo {author} {\bibfnamefont {M.}~\bibnamefont {{Baldi}}}, \bibinfo
  {author} {\bibfnamefont {K.}~\bibnamefont {{Koyama}}}, \bibinfo {author}
  {\bibfnamefont {B.}~\bibnamefont {{Li}}}, \bibinfo {author} {\bibfnamefont
  {L.}~\bibnamefont {{Lombriser}}}, \ and\ \bibinfo {author} {\bibfnamefont
  {G.-B.}\ \bibnamefont {{Zhao}}},\ }\bibfield  {title} {\enquote {\bibinfo
  {title} {{Emulators for the nonlinear matter power spectrum beyond
  {\ensuremath{\Lambda}} CDM}},}\ }\href {\doibase 10.1103/PhysRevD.100.123540}
  {\bibfield  {journal} {\bibinfo  {journal} {\prd}\ }\textbf {\bibinfo
  {volume} {100}},\ \bibinfo {eid} {123540} (\bibinfo {year} {2019})},\ \Eprint
  {http://arxiv.org/abs/1903.08798}{arXiv:1903.08798 [astro-ph.CO]}\BibitemShut
  {NoStop}%
\bibitem [{\citenamefont {{Euclid Collaboration}}\ \emph
  {et~al.}(2019)\citenamefont {{Euclid Collaboration}}, \citenamefont
  {{Knabenhans}}, \citenamefont {{Stadel}}, \citenamefont {{Marelli}},
  \citenamefont {{Potter}}, \citenamefont {{Teyssier}}, \citenamefont
  {{Legrand}}, \citenamefont {{Schneider}}, \citenamefont {{Sudret}},
  \citenamefont {{Blot}} \emph {et~al.}}]{2019MNRAS.484.5509E}%
  \BibitemOpen
  \bibfield  {author} {\bibinfo {author} {\bibnamefont {{Euclid
  Collaboration}}}, \bibinfo {author} {\bibfnamefont {M.}~\bibnamefont
  {{Knabenhans}}}, \bibinfo {author} {\bibfnamefont {J.}~\bibnamefont
  {{Stadel}}}, \bibinfo {author} {\bibfnamefont {S.}~\bibnamefont {{Marelli}}},
  \bibinfo {author} {\bibfnamefont {D.}~\bibnamefont {{Potter}}}, \bibinfo
  {author} {\bibfnamefont {R.}~\bibnamefont {{Teyssier}}}, \bibinfo {author}
  {\bibfnamefont {L.}~\bibnamefont {{Legrand}}}, \bibinfo {author}
  {\bibfnamefont {A.}~\bibnamefont {{Schneider}}}, \bibinfo {author}
  {\bibfnamefont {B.}~\bibnamefont {{Sudret}}}, \bibinfo {author}
  {\bibfnamefont {L.}~\bibnamefont {{Blot}}},  \emph {et~al.},\ }\bibfield
  {title} {\enquote {\bibinfo {title} {{Euclid preparation: II. The
  EUCLIDEMULATOR - a tool to compute the cosmology dependence of the nonlinear
  matter power spectrum}},}\ }\href {\doibase 10.1093/mnras/stz197} {\bibfield
  {journal} {\bibinfo  {journal} {\mnras}\ }\textbf {\bibinfo {volume} {484}},\
  \bibinfo {pages} {5509} (\bibinfo {year} {2019})},\ \Eprint
  {http://arxiv.org/abs/1809.04695}{arXiv:1809.04695 [astro-ph.CO]}\BibitemShut
  {NoStop}%
\bibitem [{\citenamefont {{Angulo}}\ \emph {et~al.}(2021)\citenamefont
  {{Angulo}}, \citenamefont {{Zennaro}}, \citenamefont {{Contreras}},
  \citenamefont {{Aric{\`o}}}, \citenamefont {{Pellejero-Iba{\~n}ez}},\ and\
  \citenamefont {{St{\"u}cker}}}]{2021MNRAS.507.5869A}%
  \BibitemOpen
  \bibfield  {author} {\bibinfo {author} {\bibfnamefont {R.~E.}\ \bibnamefont
  {{Angulo}}}, \bibinfo {author} {\bibfnamefont {M.}~\bibnamefont {{Zennaro}}},
  \bibinfo {author} {\bibfnamefont {S.}~\bibnamefont {{Contreras}}}, \bibinfo
  {author} {\bibfnamefont {G.}~\bibnamefont {{Aric{\`o}}}}, \bibinfo {author}
  {\bibfnamefont {M.}~\bibnamefont {{Pellejero-Iba{\~n}ez}}}, \ and\ \bibinfo
  {author} {\bibfnamefont {J.}~\bibnamefont {{St{\"u}cker}}},\ }\bibfield
  {title} {\enquote {\bibinfo {title} {{The BACCO simulation project:
  exploiting the full power of large-scale structure for cosmology}},}\ }\href
  {\doibase 10.1093/mnras/stab2018} {\bibfield  {journal} {\bibinfo  {journal}
  {\mnras}\ }\textbf {\bibinfo {volume} {507}},\ \bibinfo {pages} {5869}
  (\bibinfo {year} {2021})},\ \Eprint
  {http://arxiv.org/abs/2004.06245}{arXiv:2004.06245 [astro-ph.CO]}\BibitemShut
  {NoStop}%
\bibitem [{\citenamefont {{DESI Collaboration}}\ \emph
  {et~al.}(2016)\citenamefont {{DESI Collaboration}}, \citenamefont
  {{Aghamousa}}, \citenamefont {{Aguilar}}, \citenamefont {{Ahlen}},
  \citenamefont {{Alam}}, \citenamefont {{Allen}}, \citenamefont {{Allende
  Prieto}}, \citenamefont {{Annis}}, \citenamefont {{Bailey}}, \citenamefont
  {{Balland}} \emph {et~al.}}]{2016arXiv161100036D}%
  \BibitemOpen
  \bibfield  {author} {\bibinfo {author} {\bibnamefont {{DESI Collaboration}}},
  \bibinfo {author} {\bibfnamefont {A.}~\bibnamefont {{Aghamousa}}}, \bibinfo
  {author} {\bibfnamefont {J.}~\bibnamefont {{Aguilar}}}, \bibinfo {author}
  {\bibfnamefont {S.}~\bibnamefont {{Ahlen}}}, \bibinfo {author} {\bibfnamefont
  {S.}~\bibnamefont {{Alam}}}, \bibinfo {author} {\bibfnamefont {L.~E.}\
  \bibnamefont {{Allen}}}, \bibinfo {author} {\bibfnamefont {C.}~\bibnamefont
  {{Allende Prieto}}}, \bibinfo {author} {\bibfnamefont {J.}~\bibnamefont
  {{Annis}}}, \bibinfo {author} {\bibfnamefont {S.}~\bibnamefont {{Bailey}}},
  \bibinfo {author} {\bibfnamefont {C.}~\bibnamefont {{Balland}}},  \emph
  {et~al.},\ }\bibfield  {title} {\enquote {\bibinfo {title} {{The DESI
  Experiment Part I: Science,Targeting, and Survey Design}},}\ }\href {\doibase
  10.48550/arXiv.1611.00036} {\bibfield  {journal} {\bibinfo  {journal} {arXiv
  e-prints}\ ,\ \bibinfo {eid} {arXiv:1611.00036}} (\bibinfo {year} {2016})},\
  \Eprint {http://arxiv.org/abs/1611.00036}{arXiv:1611.00036
  [astro-ph.IM]}\BibitemShut {NoStop}%
\bibitem [{\citenamefont {{Adame}}\ \emph {et~al.}(2025)\citenamefont
  {{Adame}}, \citenamefont {{Aguilar}}, \citenamefont {{Ahlen}}, \citenamefont
  {{Alam}}, \citenamefont {{Alexander}}, \citenamefont {{Alvarez}},
  \citenamefont {{Alves}}, \citenamefont {{Anand}}, \citenamefont {{Andrade}},
  \citenamefont {{Armengaud}} \emph {et~al.}}]{2025JCAP...02..021A}%
  \BibitemOpen
  \bibfield  {author} {\bibinfo {author} {\bibfnamefont {A.~G.}\ \bibnamefont
  {{Adame}}}, \bibinfo {author} {\bibfnamefont {J.}~\bibnamefont {{Aguilar}}},
  \bibinfo {author} {\bibfnamefont {S.}~\bibnamefont {{Ahlen}}}, \bibinfo
  {author} {\bibfnamefont {S.}~\bibnamefont {{Alam}}}, \bibinfo {author}
  {\bibfnamefont {D.~M.}\ \bibnamefont {{Alexander}}}, \bibinfo {author}
  {\bibfnamefont {M.}~\bibnamefont {{Alvarez}}}, \bibinfo {author}
  {\bibfnamefont {O.}~\bibnamefont {{Alves}}}, \bibinfo {author} {\bibfnamefont
  {A.}~\bibnamefont {{Anand}}}, \bibinfo {author} {\bibfnamefont
  {U.}~\bibnamefont {{Andrade}}}, \bibinfo {author} {\bibfnamefont
  {E.}~\bibnamefont {{Armengaud}}},  \emph {et~al.},\ }\bibfield  {title}
  {\enquote {\bibinfo {title} {{DESI 2024 VI: cosmological constraints from the
  measurements of baryon acoustic oscillations}},}\ }\href {\doibase
  10.1088/1475-7516/2025/02/021} {\bibfield  {journal} {\bibinfo  {journal}
  {\jcap}\ }\textbf {\bibinfo {volume} {2025}},\ \bibinfo {eid} {021} (\bibinfo
  {year} {2025})},\ \Eprint {http://arxiv.org/abs/2404.03002}{arXiv:2404.03002
  [astro-ph.CO]}\BibitemShut {NoStop}%
\bibitem [{\citenamefont {{Laureijs}}\ \emph {et~al.}(2011)\citenamefont
  {{Laureijs}}, \citenamefont {{Amiaux}}, \citenamefont {{Arduini}},
  \citenamefont {{Augu{\`e}res}}, \citenamefont {{Brinchmann}}, \citenamefont
  {{Cole}}, \citenamefont {{Cropper}}, \citenamefont {{Dabin}}, \citenamefont
  {{Duvet}}, \citenamefont {{Ealet}} \emph {et~al.}}]{2011arXiv1110.3193L}%
  \BibitemOpen
  \bibfield  {author} {\bibinfo {author} {\bibfnamefont {R.}~\bibnamefont
  {{Laureijs}}}, \bibinfo {author} {\bibfnamefont {J.}~\bibnamefont
  {{Amiaux}}}, \bibinfo {author} {\bibfnamefont {S.}~\bibnamefont {{Arduini}}},
  \bibinfo {author} {\bibfnamefont {J.~L.}\ \bibnamefont {{Augu{\`e}res}}},
  \bibinfo {author} {\bibfnamefont {J.}~\bibnamefont {{Brinchmann}}}, \bibinfo
  {author} {\bibfnamefont {R.}~\bibnamefont {{Cole}}}, \bibinfo {author}
  {\bibfnamefont {M.}~\bibnamefont {{Cropper}}}, \bibinfo {author}
  {\bibfnamefont {C.}~\bibnamefont {{Dabin}}}, \bibinfo {author} {\bibfnamefont
  {L.}~\bibnamefont {{Duvet}}}, \bibinfo {author} {\bibfnamefont
  {A.}~\bibnamefont {{Ealet}}},  \emph {et~al.},\ }\bibfield  {title} {\enquote
  {\bibinfo {title} {{Euclid Definition Study Report}},}\ }\href {\doibase
  10.48550/arXiv.1110.3193} {\bibfield  {journal} {\bibinfo  {journal} {arXiv
  e-prints}\ ,\ \bibinfo {eid} {arXiv:1110.3193}} (\bibinfo {year} {2011})},\
  \Eprint {http://arxiv.org/abs/1110.3193}{arXiv:1110.3193
  [astro-ph.CO]}\BibitemShut {NoStop}%
\bibitem [{\citenamefont {{Euclid Collaboration}}\ \emph
  {et~al.}(2022)\citenamefont {{Euclid Collaboration}}, \citenamefont
  {{Ili{\'c}}}, \citenamefont {{Aghanim}}, \citenamefont {{Baccigalupi}},
  \citenamefont {{Bermejo-Climent}}, \citenamefont {{Fabbian}}, \citenamefont
  {{Legrand}}, \citenamefont {{Paoletti}}, \citenamefont {{Ballardini}},
  \citenamefont {{Archidiacono}} \emph {et~al.}}]{2022A&A...657A..91E}%
  \BibitemOpen
  \bibfield  {author} {\bibinfo {author} {\bibnamefont {{Euclid
  Collaboration}}}, \bibinfo {author} {\bibfnamefont {S.}~\bibnamefont
  {{Ili{\'c}}}}, \bibinfo {author} {\bibfnamefont {N.}~\bibnamefont
  {{Aghanim}}}, \bibinfo {author} {\bibfnamefont {C.}~\bibnamefont
  {{Baccigalupi}}}, \bibinfo {author} {\bibfnamefont {J.~R.}\ \bibnamefont
  {{Bermejo-Climent}}}, \bibinfo {author} {\bibfnamefont {G.}~\bibnamefont
  {{Fabbian}}}, \bibinfo {author} {\bibfnamefont {L.}~\bibnamefont
  {{Legrand}}}, \bibinfo {author} {\bibfnamefont {D.}~\bibnamefont
  {{Paoletti}}}, \bibinfo {author} {\bibfnamefont {M.}~\bibnamefont
  {{Ballardini}}}, \bibinfo {author} {\bibfnamefont {M.}~\bibnamefont
  {{Archidiacono}}},  \emph {et~al.},\ }\bibfield  {title} {\enquote {\bibinfo
  {title} {{Euclid preparation. XV. Forecasting cosmological constraints for
  the Euclid and CMB joint analysis}},}\ }\href {\doibase
  10.1051/0004-6361/202141556} {\bibfield  {journal} {\bibinfo  {journal}
  {\aap}\ }\textbf {\bibinfo {volume} {657}},\ \bibinfo {eid} {A91} (\bibinfo
  {year} {2022})},\ \Eprint {http://arxiv.org/abs/2106.08346}{arXiv:2106.08346
  [astro-ph.CO]}\BibitemShut {NoStop}%
\bibitem [{\citenamefont {{Benitez}}\ \emph {et~al.}(2014)\citenamefont
  {{Benitez}}, \citenamefont {{Dupke}}, \citenamefont {{Moles}}, \citenamefont
  {{Sodre}}, \citenamefont {{Cenarro}}, \citenamefont {{Marin-Franch}},
  \citenamefont {{Taylor}}, \citenamefont {{Cristobal}}, \citenamefont
  {{Fernandez-Soto}}, \citenamefont {{Mendes de Oliveira}} \emph
  {et~al.}}]{2014arXiv1403.5237B}%
  \BibitemOpen
  \bibfield  {author} {\bibinfo {author} {\bibfnamefont {N.}~\bibnamefont
  {{Benitez}}}, \bibinfo {author} {\bibfnamefont {R.}~\bibnamefont {{Dupke}}},
  \bibinfo {author} {\bibfnamefont {M.}~\bibnamefont {{Moles}}}, \bibinfo
  {author} {\bibfnamefont {L.}~\bibnamefont {{Sodre}}}, \bibinfo {author}
  {\bibfnamefont {J.}~\bibnamefont {{Cenarro}}}, \bibinfo {author}
  {\bibfnamefont {A.}~\bibnamefont {{Marin-Franch}}}, \bibinfo {author}
  {\bibfnamefont {K.}~\bibnamefont {{Taylor}}}, \bibinfo {author}
  {\bibfnamefont {D.}~\bibnamefont {{Cristobal}}}, \bibinfo {author}
  {\bibfnamefont {A.}~\bibnamefont {{Fernandez-Soto}}}, \bibinfo {author}
  {\bibfnamefont {C.}~\bibnamefont {{Mendes de Oliveira}}},  \emph {et~al.},\
  }\bibfield  {title} {\enquote {\bibinfo {title} {{J-PAS: The
  Javalambre-Physics of the Accelerated Universe Astrophysical Survey}},}\
  }\href {\doibase 10.48550/arXiv.1403.5237} {\bibfield  {journal} {\bibinfo
  {journal} {arXiv e-prints}\ ,\ \bibinfo {eid} {arXiv:1403.5237}} (\bibinfo
  {year} {2014})},\ \Eprint {http://arxiv.org/abs/1403.5237}{arXiv:1403.5237
  [astro-ph.CO]}\BibitemShut {NoStop}%
\bibitem [{\citenamefont {{Gong}}\ \emph {et~al.}(2025)\citenamefont {{Gong}},
  \citenamefont {{Miao}}, \citenamefont {{Zhou}}, \citenamefont {{Xiong}},
  \citenamefont {{Song}}, \citenamefont {{Jiang}}, \citenamefont {{Wang}},
  \citenamefont {{Yan}}, \citenamefont {{Wu}}, \citenamefont {{Deng}} \emph
  {et~al.}}]{2025SCPMA..6880402G}%
  \BibitemOpen
  \bibfield  {author} {\bibinfo {author} {\bibfnamefont {Y.}~\bibnamefont
  {{Gong}}}, \bibinfo {author} {\bibfnamefont {H.}~\bibnamefont {{Miao}}},
  \bibinfo {author} {\bibfnamefont {X.}~\bibnamefont {{Zhou}}}, \bibinfo
  {author} {\bibfnamefont {Q.}~\bibnamefont {{Xiong}}}, \bibinfo {author}
  {\bibfnamefont {Y.}~\bibnamefont {{Song}}}, \bibinfo {author} {\bibfnamefont
  {Y.}~\bibnamefont {{Jiang}}}, \bibinfo {author} {\bibfnamefont
  {M.}~\bibnamefont {{Wang}}}, \bibinfo {author} {\bibfnamefont
  {J.}~\bibnamefont {{Yan}}}, \bibinfo {author} {\bibfnamefont
  {B.}~\bibnamefont {{Wu}}}, \bibinfo {author} {\bibfnamefont {F.}~\bibnamefont
  {{Deng}}},  \emph {et~al.},\ }\bibfield  {title} {\enquote {\bibinfo {title}
  {{Future cosmology: New physics and opportunity from the China Space Station
  Telescope (CSST)}},}\ }\href {\doibase 10.1007/s11433-025-2646-2} {\bibfield
  {journal} {\bibinfo  {journal} {Science China Physics, Mechanics, and
  Astronomy}\ }\textbf {\bibinfo {volume} {68}},\ \bibinfo {eid} {280402}
  (\bibinfo {year} {2025})},\ \Eprint
  {http://arxiv.org/abs/2501.15023}{arXiv:2501.15023 [astro-ph.CO]}\BibitemShut
  {NoStop}%
\bibitem [{\citenamefont {{Verdier}}\ \emph {et~al.}(2025)\citenamefont
  {{Verdier}}, \citenamefont {{Rocher}}, \citenamefont {{Bandi}}, \citenamefont
  {{Richard}}, \citenamefont {{Roukema}}, \citenamefont {{Loveday}},
  \citenamefont {{Tempel}}, \citenamefont {{Bilicki}}, \citenamefont
  {{Kneib}},\ and\ \citenamefont {{Guitton}}}]{2025arXiv250807311V}%
  \BibitemOpen
  \bibfield  {author} {\bibinfo {author} {\bibfnamefont {A.}~\bibnamefont
  {{Verdier}}}, \bibinfo {author} {\bibfnamefont {A.}~\bibnamefont {{Rocher}}},
  \bibinfo {author} {\bibfnamefont {B.}~\bibnamefont {{Bandi}}}, \bibinfo
  {author} {\bibfnamefont {J.}~\bibnamefont {{Richard}}}, \bibinfo {author}
  {\bibfnamefont {B.}~\bibnamefont {{Roukema}}}, \bibinfo {author}
  {\bibfnamefont {J.}~\bibnamefont {{Loveday}}}, \bibinfo {author}
  {\bibfnamefont {E.}~\bibnamefont {{Tempel}}}, \bibinfo {author}
  {\bibfnamefont {M.}~\bibnamefont {{Bilicki}}}, \bibinfo {author}
  {\bibfnamefont {J.-P.}\ \bibnamefont {{Kneib}}}, \ and\ \bibinfo {author}
  {\bibfnamefont {M.}~\bibnamefont {{Guitton}}},\ }\bibfield  {title} {\enquote
  {\bibinfo {title} {{The 4MOST-Cosmology Redshift Survey: Target Selection of
  Bright Galaxies and Luminous Red Galaxies}},}\ }\href {\doibase
  10.48550/arXiv.2508.07311} {\bibfield  {journal} {\bibinfo  {journal} {arXiv
  e-prints}\ ,\ \bibinfo {eid} {arXiv:2508.07311}} (\bibinfo {year} {2025})},\
  \Eprint {http://arxiv.org/abs/2508.07311}{arXiv:2508.07311
  [astro-ph.CO]}\BibitemShut {NoStop}%
\bibitem [{\citenamefont {{Turner}}\ \emph {et~al.}(2023)\citenamefont
  {{Turner}}, \citenamefont {{Blake}},\ and\ \citenamefont
  {{Ruggeri}}}]{2023MNRAS.518.2436T}%
  \BibitemOpen
  \bibfield  {author} {\bibinfo {author} {\bibfnamefont {R.~J.}\ \bibnamefont
  {{Turner}}}, \bibinfo {author} {\bibfnamefont {C.}~\bibnamefont {{Blake}}}, \
  and\ \bibinfo {author} {\bibfnamefont {R.}~\bibnamefont {{Ruggeri}}},\
  }\bibfield  {title} {\enquote {\bibinfo {title} {{A local measurement of the
  growth rate from peculiar velocities and galaxy clustering correlations in
  the 6dF Galaxy Survey}},}\ }\href {\doibase 10.1093/mnras/stac3256}
  {\bibfield  {journal} {\bibinfo  {journal} {\mnras}\ }\textbf {\bibinfo
  {volume} {518}},\ \bibinfo {pages} {2436} (\bibinfo {year} {2023})},\ \Eprint
  {http://arxiv.org/abs/2207.03707}{arXiv:2207.03707 [astro-ph.CO]}\BibitemShut
  {NoStop}%
\bibitem [{\citenamefont {{Lyall}}\ \emph {et~al.}(2024)\citenamefont
  {{Lyall}}, \citenamefont {{Blake}},\ and\ \citenamefont
  {{Turner}}}]{2024MNRAS.532.3972L}%
  \BibitemOpen
  \bibfield  {author} {\bibinfo {author} {\bibfnamefont {S.}~\bibnamefont
  {{Lyall}}}, \bibinfo {author} {\bibfnamefont {C.}~\bibnamefont {{Blake}}}, \
  and\ \bibinfo {author} {\bibfnamefont {R.~J.}\ \bibnamefont {{Turner}}},\
  }\bibfield  {title} {\enquote {\bibinfo {title} {{Constraining modified
  gravity scenarios with the 6dFGS and SDSS galaxy peculiar velocity data
  sets}},}\ }\href {\doibase 10.1093/mnras/stae1718} {\bibfield  {journal}
  {\bibinfo  {journal} {\mnras}\ }\textbf {\bibinfo {volume} {532}},\ \bibinfo
  {pages} {3972} (\bibinfo {year} {2024})},\ \Eprint
  {http://arxiv.org/abs/2407.18684}{arXiv:2407.18684 [astro-ph.CO]}\BibitemShut
  {NoStop}%
\bibitem [{\citenamefont {{Ferreira}}\ \emph {et~al.}(1999)\citenamefont
  {{Ferreira}}, \citenamefont {{Juszkiewicz}}, \citenamefont {{Feldman}},
  \citenamefont {{Davis}},\ and\ \citenamefont
  {{Jaffe}}}]{1999ApJ...515L...1F}%
  \BibitemOpen
  \bibfield  {author} {\bibinfo {author} {\bibfnamefont {P.~G.}\ \bibnamefont
  {{Ferreira}}}, \bibinfo {author} {\bibfnamefont {R.}~\bibnamefont
  {{Juszkiewicz}}}, \bibinfo {author} {\bibfnamefont {H.~A.}\ \bibnamefont
  {{Feldman}}}, \bibinfo {author} {\bibfnamefont {M.}~\bibnamefont {{Davis}}},
  \ and\ \bibinfo {author} {\bibfnamefont {A.~H.}\ \bibnamefont {{Jaffe}}},\
  }\bibfield  {title} {\enquote {\bibinfo {title} {{Streaming Velocities as a
  Dynamical Estimator of {\ensuremath{\Omega}}}},}\ }\href {\doibase
  10.1086/311959} {\bibfield  {journal} {\bibinfo  {journal} {\apjl}\ }\textbf
  {\bibinfo {volume} {515}},\ \bibinfo {pages} {L1} (\bibinfo {year} {1999})},\
  \Eprint {http://arxiv.org/abs/astro-ph/9812456}{arXiv:astro-ph/9812456
  [astro-ph]}\BibitemShut {NoStop}%
\bibitem [{\citenamefont {{Zhang}}\ \emph {et~al.}(2025)\citenamefont
  {{Zhang}}, \citenamefont {{Chu}}, \citenamefont {{Liao}}, \citenamefont
  {{Yeung}},\ and\ \citenamefont {{Hu}}}]{2025ApJ...978L...6Z}%
  \BibitemOpen
  \bibfield  {author} {\bibinfo {author} {\bibfnamefont {W.}~\bibnamefont
  {{Zhang}}}, \bibinfo {author} {\bibfnamefont {M.-c.}\ \bibnamefont {{Chu}}},
  \bibinfo {author} {\bibfnamefont {S.}~\bibnamefont {{Liao}}}, \bibinfo
  {author} {\bibfnamefont {S.}~\bibnamefont {{Yeung}}}, \ and\ \bibinfo
  {author} {\bibfnamefont {H.-J.}\ \bibnamefont {{Hu}}},\ }\bibfield  {title}
  {\enquote {\bibinfo {title} {{Measuring the Hubble Constant through the
  Galaxy Pairwise Peculiar Velocity}},}\ }\href {\doibase
  10.3847/2041-8213/ad9aa7} {\bibfield  {journal} {\bibinfo  {journal} {\apjl}\
  }\textbf {\bibinfo {volume} {978}},\ \bibinfo {eid} {L6} (\bibinfo {year}
  {2025})},\ \Eprint {http://arxiv.org/abs/2412.04660}{arXiv:2412.04660
  [astro-ph.CO]}\BibitemShut {NoStop}%
\bibitem [{\citenamefont {{Soergel}}\ \emph {et~al.}(2016)\citenamefont
  {{Soergel}}, \citenamefont {{Flender}}, \citenamefont {{Story}},
  \citenamefont {{Bleem}}, \citenamefont {{Giannantonio}}, \citenamefont
  {{Efstathiou}}, \citenamefont {{Rykoff}}, \citenamefont {{Benson}},
  \citenamefont {{Crawford}}, \citenamefont {{Dodelson}} \emph
  {et~al.}}]{2016MNRAS.461.3172S}%
  \BibitemOpen
  \bibfield  {author} {\bibinfo {author} {\bibfnamefont {B.}~\bibnamefont
  {{Soergel}}}, \bibinfo {author} {\bibfnamefont {S.}~\bibnamefont
  {{Flender}}}, \bibinfo {author} {\bibfnamefont {K.~T.}\ \bibnamefont
  {{Story}}}, \bibinfo {author} {\bibfnamefont {L.}~\bibnamefont {{Bleem}}},
  \bibinfo {author} {\bibfnamefont {T.}~\bibnamefont {{Giannantonio}}},
  \bibinfo {author} {\bibfnamefont {G.}~\bibnamefont {{Efstathiou}}}, \bibinfo
  {author} {\bibfnamefont {E.}~\bibnamefont {{Rykoff}}}, \bibinfo {author}
  {\bibfnamefont {B.~A.}\ \bibnamefont {{Benson}}}, \bibinfo {author}
  {\bibfnamefont {T.}~\bibnamefont {{Crawford}}}, \bibinfo {author}
  {\bibfnamefont {S.}~\bibnamefont {{Dodelson}}},  \emph {et~al.},\ }\bibfield
  {title} {\enquote {\bibinfo {title} {{Detection of the kinematic
  Sunyaev-Zel'dovich effect with DES Year 1 and SPT}},}\ }\href {\doibase
  10.1093/mnras/stw1455} {\bibfield  {journal} {\bibinfo  {journal} {\mnras}\
  }\textbf {\bibinfo {volume} {461}},\ \bibinfo {pages} {3172} (\bibinfo {year}
  {2016})},\ \Eprint {http://arxiv.org/abs/1603.03904}{arXiv:1603.03904
  [astro-ph.CO]}\BibitemShut {NoStop}%
\bibitem [{\citenamefont {{Maleubre}}\ \emph {et~al.}(2023)\citenamefont
  {{Maleubre}}, \citenamefont {{Eisenstein}}, \citenamefont {{Garrison}},\ and\
  \citenamefont {{Joyce}}}]{2023MNRAS.525.1039M}%
  \BibitemOpen
  \bibfield  {author} {\bibinfo {author} {\bibfnamefont {S.}~\bibnamefont
  {{Maleubre}}}, \bibinfo {author} {\bibfnamefont {D.~J.}\ \bibnamefont
  {{Eisenstein}}}, \bibinfo {author} {\bibfnamefont {L.~H.}\ \bibnamefont
  {{Garrison}}}, \ and\ \bibinfo {author} {\bibfnamefont {M.}~\bibnamefont
  {{Joyce}}},\ }\bibfield  {title} {\enquote {\bibinfo {title} {{Constraining
  accuracy of the pairwise velocities in N-body simulations using scale-free
  models}},}\ }\href {\doibase 10.1093/mnras/stad2388} {\bibfield  {journal}
  {\bibinfo  {journal} {\mnras}\ }\textbf {\bibinfo {volume} {525}},\ \bibinfo
  {pages} {1039} (\bibinfo {year} {2023})},\ \Eprint
  {http://arxiv.org/abs/2211.07607}{arXiv:2211.07607 [astro-ph.CO]}\BibitemShut
  {NoStop}%
\bibitem [{\citenamefont {{Xu}}(2022)}]{2022arXiv220206515X}%
  \BibitemOpen
  \bibfield  {author} {\bibinfo {author} {\bibfnamefont {Z.}~\bibnamefont
  {{Xu}}},\ }\bibfield  {title} {\enquote {\bibinfo {title} {{On the
  statistical theory of self-gravitating collisionless dark matter flow: Scale
  and redshift variation of velocity and density distributions}},}\ }\href
  {\doibase 10.48550/arXiv.2202.06515} {\bibfield  {journal} {\bibinfo
  {journal} {arXiv e-prints}\ ,\ \bibinfo {eid} {arXiv:2202.06515}} (\bibinfo
  {year} {2022})},\ \Eprint {http://arxiv.org/abs/2202.06515}{arXiv:2202.06515
  [astro-ph.CO]}\BibitemShut {NoStop}%
\bibitem [{\citenamefont {{Jaber}}\ \emph {et~al.}(2024)\citenamefont
  {{Jaber}}, \citenamefont {{Hellwing}}, \citenamefont {{Garc{\'\i}a-Farieta}},
  \citenamefont {{Gupta}},\ and\ \citenamefont
  {{Bilicki}}}]{2024PhRvD.109l3528J}%
  \BibitemOpen
  \bibfield  {author} {\bibinfo {author} {\bibfnamefont {M.}~\bibnamefont
  {{Jaber}}}, \bibinfo {author} {\bibfnamefont {W.~A.}\ \bibnamefont
  {{Hellwing}}}, \bibinfo {author} {\bibfnamefont {J.~E.}\ \bibnamefont
  {{Garc{\'\i}a-Farieta}}}, \bibinfo {author} {\bibfnamefont {S.}~\bibnamefont
  {{Gupta}}}, \ and\ \bibinfo {author} {\bibfnamefont {M.}~\bibnamefont
  {{Bilicki}}},\ }\bibfield  {title} {\enquote {\bibinfo {title} {{Dynamics of
  pairwise motions in the fully nonlinear regime in LCDM and modified gravity
  cosmologies}},}\ }\href {\doibase 10.1103/PhysRevD.109.123528} {\bibfield
  {journal} {\bibinfo  {journal} {\prd}\ }\textbf {\bibinfo {volume} {109}},\
  \bibinfo {eid} {123528} (\bibinfo {year} {2024})},\ \Eprint
  {http://arxiv.org/abs/2312.00472}{arXiv:2312.00472 [astro-ph.CO]}\BibitemShut
  {NoStop}%
\bibitem [{\citenamefont {{Bibiano}}\ and\ \citenamefont
  {{Croton}}(2017)}]{2017MNRAS.467.1386B}%
  \BibitemOpen
  \bibfield  {author} {\bibinfo {author} {\bibfnamefont {A.}~\bibnamefont
  {{Bibiano}}}\ and\ \bibinfo {author} {\bibfnamefont {D.~J.}\ \bibnamefont
  {{Croton}}},\ }\bibfield  {title} {\enquote {\bibinfo {title} {{Pairwise
  velocities in the ``Running FLRW'' cosmological model}},}\ }\href {\doibase
  10.1093/mnras/stx070} {\bibfield  {journal} {\bibinfo  {journal} {\mnras}\
  }\textbf {\bibinfo {volume} {467}},\ \bibinfo {pages} {1386} (\bibinfo {year}
  {2017})},\ \Eprint {http://arxiv.org/abs/1701.04453}{arXiv:1701.04453
  [astro-ph.CO]}\BibitemShut {NoStop}%
\bibitem [{\citenamefont {{Zhang}}\ \emph {et~al.}(2024)\citenamefont
  {{Zhang}}, \citenamefont {{Chu}}, \citenamefont {{Hu}}, \citenamefont
  {{Liao}},\ and\ \citenamefont {{Yeung}}}]{2024MNRAS.529..360Z}%
  \BibitemOpen
  \bibfield  {author} {\bibinfo {author} {\bibfnamefont {W.}~\bibnamefont
  {{Zhang}}}, \bibinfo {author} {\bibfnamefont {M.-c.}\ \bibnamefont {{Chu}}},
  \bibinfo {author} {\bibfnamefont {R.}~\bibnamefont {{Hu}}}, \bibinfo {author}
  {\bibfnamefont {S.}~\bibnamefont {{Liao}}}, \ and\ \bibinfo {author}
  {\bibfnamefont {S.}~\bibnamefont {{Yeung}}},\ }\bibfield  {title} {\enquote
  {\bibinfo {title} {{Measuring neutrino mass and asymmetry with matter
  pairwise velocities}},}\ }\href {\doibase 10.1093/mnras/stae511} {\bibfield
  {journal} {\bibinfo  {journal} {\mnras}\ }\textbf {\bibinfo {volume} {529}},\
  \bibinfo {pages} {360} (\bibinfo {year} {2024})},\ \Eprint
  {http://arxiv.org/abs/2312.04278}{arXiv:2312.04278 [astro-ph.CO]}\BibitemShut
  {NoStop}%
\bibitem [{\citenamefont {{Peebles}}(1976)}]{1976Ap&SS..45....3P}%
  \BibitemOpen
  \bibfield  {author} {\bibinfo {author} {\bibfnamefont {P.~J.~E.}\
  \bibnamefont {{Peebles}}},\ }\bibfield  {title} {\enquote {\bibinfo {title}
  {{A Cosmic Virial Theorem}},}\ }\href {\doibase 10.1007/BF00642136}
  {\bibfield  {journal} {\bibinfo  {journal} {\apss}\ }\textbf {\bibinfo
  {volume} {45}},\ \bibinfo {pages} {3} (\bibinfo {year} {1976})}\BibitemShut
  {NoStop}%
\bibitem [{\citenamefont {{Davis}}\ and\ \citenamefont
  {{Peebles}}(1983)}]{1983ApJ...267..465D}%
  \BibitemOpen
  \bibfield  {author} {\bibinfo {author} {\bibfnamefont {M.}~\bibnamefont
  {{Davis}}}\ and\ \bibinfo {author} {\bibfnamefont {P.~J.~E.}\ \bibnamefont
  {{Peebles}}},\ }\bibfield  {title} {\enquote {\bibinfo {title} {{A survey of
  galaxy redshifts. V. The two-point position and velocity correlations.}}}\
  }\href {\doibase 10.1086/160884} {\bibfield  {journal} {\bibinfo  {journal}
  {\apj}\ }\textbf {\bibinfo {volume} {267}},\ \bibinfo {pages} {465} (\bibinfo
  {year} {1983})}\BibitemShut {NoStop}%
\bibitem [{\citenamefont {Juszkiewicz}\ \emph {et~al.}(1999)\citenamefont
  {Juszkiewicz}, \citenamefont {Springel},\ and\ \citenamefont
  {Durrer}}]{Juszkiewicz_1999}%
  \BibitemOpen
  \bibfield  {author} {\bibinfo {author} {\bibfnamefont {R.}~\bibnamefont
  {Juszkiewicz}}, \bibinfo {author} {\bibfnamefont {V.}~\bibnamefont
  {Springel}}, \ and\ \bibinfo {author} {\bibfnamefont {R.}~\bibnamefont
  {Durrer}},\ }\bibfield  {title} {\enquote {\bibinfo {title} {Dynamics of
  pairwise motions},}\ }\href {\doibase 10.1086/312055} {\bibfield  {journal}
  {\bibinfo  {journal} {The Astrophysical Journal}\ }\textbf {\bibinfo {volume}
  {518}},\ \bibinfo {pages} {L25} (\bibinfo {year} {1999})}\BibitemShut
  {NoStop}%
\bibitem [{\citenamefont {{Peebles}}(1980)}]{1980lssu.book}%
  \BibitemOpen
  \bibfield  {author} {\bibinfo {author} {\bibfnamefont {P.~J.~E.}\
  \bibnamefont {{Peebles}}},\ }\href@noop {} {\emph {\bibinfo {title} {{The
  large-scale structure of the universe}}}}\ (\bibinfo  {publisher} {Princeton
  University Press},\ \bibinfo {year} {1980})\BibitemShut {NoStop}%
\bibitem [{\citenamefont {{Huterer}}(2023)}]{2023A&ARv..31....2H}%
  \BibitemOpen
  \bibfield  {author} {\bibinfo {author} {\bibfnamefont {D.}~\bibnamefont
  {{Huterer}}},\ }\bibfield  {title} {\enquote {\bibinfo {title} {{Growth of
  cosmic structure}},}\ }\href {\doibase 10.1007/s00159-023-00147-4} {\bibfield
   {journal} {\bibinfo  {journal} {\aapr}\ }\textbf {\bibinfo {volume} {31}},\
  \bibinfo {eid} {2} (\bibinfo {year} {2023})},\ \Eprint
  {http://arxiv.org/abs/2212.05003}{arXiv:2212.05003 [astro-ph.CO]}\BibitemShut
  {NoStop}%
\bibitem [{\citenamefont {{Carlson}}\ \emph {et~al.}(2013)\citenamefont
  {{Carlson}}, \citenamefont {{Reid}},\ and\ \citenamefont
  {{White}}}]{2013MNRAS.429.1674C}%
  \BibitemOpen
  \bibfield  {author} {\bibinfo {author} {\bibfnamefont {J.}~\bibnamefont
  {{Carlson}}}, \bibinfo {author} {\bibfnamefont {B.}~\bibnamefont {{Reid}}}, \
  and\ \bibinfo {author} {\bibfnamefont {M.}~\bibnamefont {{White}}},\
  }\bibfield  {title} {\enquote {\bibinfo {title} {{Convolution Lagrangian
  perturbation theory for biased tracers}},}\ }\href {\doibase
  10.1093/mnras/sts457} {\bibfield  {journal} {\bibinfo  {journal} {\mnras}\
  }\textbf {\bibinfo {volume} {429}},\ \bibinfo {pages} {1674} (\bibinfo {year}
  {2013})},\ \Eprint {http://arxiv.org/abs/1209.0780}{arXiv:1209.0780
  [astro-ph.CO]}\BibitemShut {NoStop}%
\bibitem [{\citenamefont {{Carrasco}}\ \emph {et~al.}(2014)\citenamefont
  {{Carrasco}}, \citenamefont {{Foreman}}, \citenamefont {{Green}},\ and\
  \citenamefont {{Senatore}}}]{2014JCAP...07..057C}%
  \BibitemOpen
  \bibfield  {author} {\bibinfo {author} {\bibfnamefont {J.~J.~M.}\
  \bibnamefont {{Carrasco}}}, \bibinfo {author} {\bibfnamefont
  {S.}~\bibnamefont {{Foreman}}}, \bibinfo {author} {\bibfnamefont
  {D.}~\bibnamefont {{Green}}}, \ and\ \bibinfo {author} {\bibfnamefont
  {L.}~\bibnamefont {{Senatore}}},\ }\bibfield  {title} {\enquote {\bibinfo
  {title} {{The Effective Field Theory of Large Scale Structures at two
  loops}},}\ }\href {\doibase 10.1088/1475-7516/2014/07/057} {\bibfield
  {journal} {\bibinfo  {journal} {\jcap}\ }\textbf {\bibinfo {volume} {2014}},\
  \bibinfo {eid} {057} (\bibinfo {year} {2014})},\ \Eprint
  {http://arxiv.org/abs/1310.0464}{arXiv:1310.0464 [astro-ph.CO]}\BibitemShut
  {NoStop}%
\bibitem [{\citenamefont {{Baldauf}}\ \emph {et~al.}(2015)\citenamefont
  {{Baldauf}}, \citenamefont {{Mercolli}},\ and\ \citenamefont
  {{Zaldarriaga}}}]{2015PhRvD..92l3007B}%
  \BibitemOpen
  \bibfield  {author} {\bibinfo {author} {\bibfnamefont {T.}~\bibnamefont
  {{Baldauf}}}, \bibinfo {author} {\bibfnamefont {L.}~\bibnamefont
  {{Mercolli}}}, \ and\ \bibinfo {author} {\bibfnamefont {M.}~\bibnamefont
  {{Zaldarriaga}}},\ }\bibfield  {title} {\enquote {\bibinfo {title}
  {{Effective field theory of large scale structure at two loops: The apparent
  scale dependence of the speed of sound}},}\ }\href {\doibase
  10.1103/PhysRevD.92.123007} {\bibfield  {journal} {\bibinfo  {journal}
  {\prd}\ }\textbf {\bibinfo {volume} {92}},\ \bibinfo {eid} {123007} (\bibinfo
  {year} {2015})},\ \Eprint {http://arxiv.org/abs/1507.02256}{arXiv:1507.02256
  [astro-ph.CO]}\BibitemShut {NoStop}%
\bibitem [{\citenamefont {{Ivanov}}(2022)}]{2022arXiv221208488I}%
  \BibitemOpen
  \bibfield  {author} {\bibinfo {author} {\bibfnamefont {M.~M.}\ \bibnamefont
  {{Ivanov}}},\ }\bibfield  {title} {\enquote {\bibinfo {title} {{Effective
  Field Theory for Large Scale Structure}},}\ }\href {\doibase
  10.48550/arXiv.2212.08488} {\bibfield  {journal} {\bibinfo  {journal} {arXiv
  e-prints}\ ,\ \bibinfo {eid} {arXiv:2212.08488}} (\bibinfo {year} {2022})},\
  \Eprint {http://arxiv.org/abs/2212.08488}{arXiv:2212.08488
  [astro-ph.CO]}\BibitemShut {NoStop}%
\bibitem [{\citenamefont {{Blas}}\ \emph {et~al.}(2011)\citenamefont {{Blas}},
  \citenamefont {{Lesgourgues}},\ and\ \citenamefont
  {{Tram}}}]{2011JCAP...07..034B}%
  \BibitemOpen
  \bibfield  {author} {\bibinfo {author} {\bibfnamefont {D.}~\bibnamefont
  {{Blas}}}, \bibinfo {author} {\bibfnamefont {J.}~\bibnamefont
  {{Lesgourgues}}}, \ and\ \bibinfo {author} {\bibfnamefont {T.}~\bibnamefont
  {{Tram}}},\ }\bibfield  {title} {\enquote {\bibinfo {title} {{The Cosmic
  Linear Anisotropy Solving System (CLASS). Part II: Approximation schemes}},}\
  }\href {\doibase 10.1088/1475-7516/2011/07/034} {\bibfield  {journal}
  {\bibinfo  {journal} {\jcap}\ }\textbf {\bibinfo {volume} {2011}},\ \bibinfo
  {eid} {034} (\bibinfo {year} {2011})},\ \Eprint
  {http://arxiv.org/abs/1104.2933}{arXiv:1104.2933 [astro-ph.CO]}\BibitemShut
  {NoStop}%
\bibitem [{\citenamefont {{Casarini}}\ \emph {et~al.}(2009)\citenamefont
  {{Casarini}}, \citenamefont {{Macci{\`o}}},\ and\ \citenamefont
  {{Bonometto}}}]{2009JCAP...03..014C}%
  \BibitemOpen
  \bibfield  {author} {\bibinfo {author} {\bibfnamefont {L.}~\bibnamefont
  {{Casarini}}}, \bibinfo {author} {\bibfnamefont {A.~V.}\ \bibnamefont
  {{Macci{\`o}}}}, \ and\ \bibinfo {author} {\bibfnamefont {S.~A.}\
  \bibnamefont {{Bonometto}}},\ }\bibfield  {title} {\enquote {\bibinfo {title}
  {{Dynamical dark energy simulations: high accuracy power spectra at high
  redshift}},}\ }\href {\doibase 10.1088/1475-7516/2009/03/014} {\bibfield
  {journal} {\bibinfo  {journal} {\jcap}\ }\textbf {\bibinfo {volume} {2009}},\
  \bibinfo {eid} {014} (\bibinfo {year} {2009})},\ \Eprint
  {http://arxiv.org/abs/0810.0190}{arXiv:0810.0190 [astro-ph]}\BibitemShut
  {NoStop}%
\bibitem [{\citenamefont {{Bird}}\ \emph {et~al.}(2012)\citenamefont {{Bird}},
  \citenamefont {{Viel}},\ and\ \citenamefont
  {{Haehnelt}}}]{2012MNRAS.420.2551B}%
  \BibitemOpen
  \bibfield  {author} {\bibinfo {author} {\bibfnamefont {S.}~\bibnamefont
  {{Bird}}}, \bibinfo {author} {\bibfnamefont {M.}~\bibnamefont {{Viel}}}, \
  and\ \bibinfo {author} {\bibfnamefont {M.~G.}\ \bibnamefont {{Haehnelt}}},\
  }\bibfield  {title} {\enquote {\bibinfo {title} {{Massive neutrinos and the
  non-linear matter power spectrum}},}\ }\href {\doibase
  10.1111/j.1365-2966.2011.20222.x} {\bibfield  {journal} {\bibinfo  {journal}
  {\mnras}\ }\textbf {\bibinfo {volume} {420}},\ \bibinfo {pages} {2551}
  (\bibinfo {year} {2012})},\ \Eprint
  {http://arxiv.org/abs/1109.4416}{arXiv:1109.4416 [astro-ph.CO]}\BibitemShut
  {NoStop}%
\bibitem [{\citenamefont {{Parimbelli}}\ \emph {et~al.}(2022)\citenamefont
  {{Parimbelli}}, \citenamefont {{Carbone}}, \citenamefont {{Bel}},
  \citenamefont {{Bose}}, \citenamefont {{Calabrese}}, \citenamefont
  {{Carella}},\ and\ \citenamefont {{Zennaro}}}]{2022JCAP...11..041P}%
  \BibitemOpen
  \bibfield  {author} {\bibinfo {author} {\bibfnamefont {G.}~\bibnamefont
  {{Parimbelli}}}, \bibinfo {author} {\bibfnamefont {C.}~\bibnamefont
  {{Carbone}}}, \bibinfo {author} {\bibfnamefont {J.}~\bibnamefont {{Bel}}},
  \bibinfo {author} {\bibfnamefont {B.}~\bibnamefont {{Bose}}}, \bibinfo
  {author} {\bibfnamefont {M.}~\bibnamefont {{Calabrese}}}, \bibinfo {author}
  {\bibfnamefont {E.}~\bibnamefont {{Carella}}}, \ and\ \bibinfo {author}
  {\bibfnamefont {M.}~\bibnamefont {{Zennaro}}},\ }\bibfield  {title} {\enquote
  {\bibinfo {title} {{DEMNUni: comparing nonlinear power spectra prescriptions
  in the presence of massive neutrinos and dynamical dark energy}},}\ }\href
  {\doibase 10.1088/1475-7516/2022/11/041} {\bibfield  {journal} {\bibinfo
  {journal} {\jcap}\ }\textbf {\bibinfo {volume} {2022}},\ \bibinfo {eid} {041}
  (\bibinfo {year} {2022})},\ \Eprint
  {http://arxiv.org/abs/2207.13677}{arXiv:2207.13677 [astro-ph.CO]}\BibitemShut
  {NoStop}%
\bibitem [{\citenamefont {{Mead}}\ \emph {et~al.}(2021)\citenamefont {{Mead}},
  \citenamefont {{Brieden}}, \citenamefont {{Tr{\"o}ster}},\ and\ \citenamefont
  {{Heymans}}}]{2021MNRAS.502.1401M}%
  \BibitemOpen
  \bibfield  {author} {\bibinfo {author} {\bibfnamefont {A.~J.}\ \bibnamefont
  {{Mead}}}, \bibinfo {author} {\bibfnamefont {S.}~\bibnamefont {{Brieden}}},
  \bibinfo {author} {\bibfnamefont {T.}~\bibnamefont {{Tr{\"o}ster}}}, \ and\
  \bibinfo {author} {\bibfnamefont {C.}~\bibnamefont {{Heymans}}},\ }\bibfield
  {title} {\enquote {\bibinfo {title} {{HMCODE-2020: improved modelling of
  non-linear cosmological power spectra with baryonic feedback}},}\ }\href
  {\doibase 10.1093/mnras/stab082} {\bibfield  {journal} {\bibinfo  {journal}
  {\mnras}\ }\textbf {\bibinfo {volume} {502}},\ \bibinfo {pages} {1401}
  (\bibinfo {year} {2021})},\ \Eprint
  {http://arxiv.org/abs/2009.01858}{arXiv:2009.01858 [astro-ph.CO]}\BibitemShut
  {NoStop}%
\bibitem [{\citenamefont {{Lawrence}}\ \emph {et~al.}(2017)\citenamefont
  {{Lawrence}}, \citenamefont {{Heitmann}}, \citenamefont {{Kwan}},
  \citenamefont {{Upadhye}}, \citenamefont {{Bingham}}, \citenamefont
  {{Habib}}, \citenamefont {{Higdon}}, \citenamefont {{Pope}}, \citenamefont
  {{Finkel}},\ and\ \citenamefont {{Frontiere}}}]{2017ApJ...847...50L}%
  \BibitemOpen
  \bibfield  {author} {\bibinfo {author} {\bibfnamefont {E.}~\bibnamefont
  {{Lawrence}}}, \bibinfo {author} {\bibfnamefont {K.}~\bibnamefont
  {{Heitmann}}}, \bibinfo {author} {\bibfnamefont {J.}~\bibnamefont {{Kwan}}},
  \bibinfo {author} {\bibfnamefont {A.}~\bibnamefont {{Upadhye}}}, \bibinfo
  {author} {\bibfnamefont {D.}~\bibnamefont {{Bingham}}}, \bibinfo {author}
  {\bibfnamefont {S.}~\bibnamefont {{Habib}}}, \bibinfo {author} {\bibfnamefont
  {D.}~\bibnamefont {{Higdon}}}, \bibinfo {author} {\bibfnamefont
  {A.}~\bibnamefont {{Pope}}}, \bibinfo {author} {\bibfnamefont
  {H.}~\bibnamefont {{Finkel}}}, \ and\ \bibinfo {author} {\bibfnamefont
  {N.}~\bibnamefont {{Frontiere}}},\ }\bibfield  {title} {\enquote {\bibinfo
  {title} {{The Mira-Titan Universe. II. Matter Power Spectrum Emulation}},}\
  }\href {\doibase 10.3847/1538-4357/aa86a9} {\bibfield  {journal} {\bibinfo
  {journal} {\apj}\ }\textbf {\bibinfo {volume} {847}},\ \bibinfo {eid} {50}
  (\bibinfo {year} {2017})},\ \Eprint
  {http://arxiv.org/abs/1705.03388}{arXiv:1705.03388 [astro-ph.CO]}\BibitemShut
  {NoStop}%
\bibitem [{\citenamefont {{Moran}}\ \emph {et~al.}(2023)\citenamefont
  {{Moran}}, \citenamefont {{Heitmann}}, \citenamefont {{Lawrence}},
  \citenamefont {{Habib}}, \citenamefont {{Bingham}}, \citenamefont
  {{Upadhye}}, \citenamefont {{Kwan}}, \citenamefont {{Higdon}},\ and\
  \citenamefont {{Payne}}}]{2023MNRAS.520.3443M}%
  \BibitemOpen
  \bibfield  {author} {\bibinfo {author} {\bibfnamefont {K.~R.}\ \bibnamefont
  {{Moran}}}, \bibinfo {author} {\bibfnamefont {K.}~\bibnamefont {{Heitmann}}},
  \bibinfo {author} {\bibfnamefont {E.}~\bibnamefont {{Lawrence}}}, \bibinfo
  {author} {\bibfnamefont {S.}~\bibnamefont {{Habib}}}, \bibinfo {author}
  {\bibfnamefont {D.}~\bibnamefont {{Bingham}}}, \bibinfo {author}
  {\bibfnamefont {A.}~\bibnamefont {{Upadhye}}}, \bibinfo {author}
  {\bibfnamefont {J.}~\bibnamefont {{Kwan}}}, \bibinfo {author} {\bibfnamefont
  {D.}~\bibnamefont {{Higdon}}}, \ and\ \bibinfo {author} {\bibfnamefont
  {R.}~\bibnamefont {{Payne}}},\ }\bibfield  {title} {\enquote {\bibinfo
  {title} {{The Mira-Titan Universe - IV. High-precision power spectrum
  emulation}},}\ }\href {\doibase 10.1093/mnras/stac3452} {\bibfield  {journal}
  {\bibinfo  {journal} {\mnras}\ }\textbf {\bibinfo {volume} {520}},\ \bibinfo
  {pages} {3443} (\bibinfo {year} {2023})},\ \Eprint
  {http://arxiv.org/abs/2207.12345}{arXiv:2207.12345 [astro-ph.CO]}\BibitemShut
  {NoStop}%
\bibitem [{\citenamefont {{Heitmann}}\ \emph {et~al.}(2016)\citenamefont
  {{Heitmann}}, \citenamefont {{Bingham}}, \citenamefont {{Lawrence}},
  \citenamefont {{Bergner}}, \citenamefont {{Habib}}, \citenamefont {{Higdon}},
  \citenamefont {{Pope}}, \citenamefont {{Biswas}}, \citenamefont {{Finkel}},
  \citenamefont {{Frontiere}},\ and\ \citenamefont
  {{Bhattacharya}}}]{2016ApJ...820..108H}%
  \BibitemOpen
  \bibfield  {author} {\bibinfo {author} {\bibfnamefont {K.}~\bibnamefont
  {{Heitmann}}}, \bibinfo {author} {\bibfnamefont {D.}~\bibnamefont
  {{Bingham}}}, \bibinfo {author} {\bibfnamefont {E.}~\bibnamefont
  {{Lawrence}}}, \bibinfo {author} {\bibfnamefont {S.}~\bibnamefont
  {{Bergner}}}, \bibinfo {author} {\bibfnamefont {S.}~\bibnamefont {{Habib}}},
  \bibinfo {author} {\bibfnamefont {D.}~\bibnamefont {{Higdon}}}, \bibinfo
  {author} {\bibfnamefont {A.}~\bibnamefont {{Pope}}}, \bibinfo {author}
  {\bibfnamefont {R.}~\bibnamefont {{Biswas}}}, \bibinfo {author}
  {\bibfnamefont {H.}~\bibnamefont {{Finkel}}}, \bibinfo {author}
  {\bibfnamefont {N.}~\bibnamefont {{Frontiere}}}, \ and\ \bibinfo {author}
  {\bibfnamefont {S.}~\bibnamefont {{Bhattacharya}}},\ }\bibfield  {title}
  {\enquote {\bibinfo {title} {{The Mira-Titan Universe: Precision Predictions
  for Dark Energy Surveys}},}\ }\href {\doibase 10.3847/0004-637X/820/2/108}
  {\bibfield  {journal} {\bibinfo  {journal} {\apj}\ }\textbf {\bibinfo
  {volume} {820}},\ \bibinfo {eid} {108} (\bibinfo {year} {2016})},\ \Eprint
  {http://arxiv.org/abs/1508.02654}{arXiv:1508.02654 [astro-ph.CO]}\BibitemShut
  {NoStop}%
\bibitem [{\citenamefont {{Chen}}\ \emph {et~al.}(2025)\citenamefont {{Chen}},
  \citenamefont {{Yu}}, \citenamefont {{Han}},\ and\ \citenamefont
  {{Jing}}}]{2025SCPMA..6889512C}%
  \BibitemOpen
  \bibfield  {author} {\bibinfo {author} {\bibfnamefont {Z.}~\bibnamefont
  {{Chen}}}, \bibinfo {author} {\bibfnamefont {Y.}~\bibnamefont {{Yu}}},
  \bibinfo {author} {\bibfnamefont {J.}~\bibnamefont {{Han}}}, \ and\ \bibinfo
  {author} {\bibfnamefont {Y.}~\bibnamefont {{Jing}}},\ }\bibfield  {title}
  {\enquote {\bibinfo {title} {{CSST cosmological emulator I: Matter power
  spectrum emulation with one percent accuracy to k = 10h Mpc$^{‑1}$}},}\
  }\href {\doibase 10.1007/s11433-025-2671-0} {\bibfield  {journal} {\bibinfo
  {journal} {Science China Physics, Mechanics, and Astronomy}\ }\textbf
  {\bibinfo {volume} {68}},\ \bibinfo {eid} {289512} (\bibinfo {year}
  {2025})},\ \Eprint {http://arxiv.org/abs/2502.11160}{arXiv:2502.11160
  [astro-ph.CO]}\BibitemShut {NoStop}%
\bibitem [{\citenamefont {{CSST Collaboration}}\ \emph
  {et~al.}(2025)\citenamefont {{CSST Collaboration}}, \citenamefont {{Gong}},
  \citenamefont {{Miao}}, \citenamefont {{Zhan}}, \citenamefont {{Li}},
  \citenamefont {{Shangguan}}, \citenamefont {{Li}}, \citenamefont {{Liu}},
  \citenamefont {{Chen}}, \citenamefont {{Yuan}} \emph
  {et~al.}}]{2025arXiv250704618C}%
  \BibitemOpen
  \bibfield  {author} {\bibinfo {author} {\bibnamefont {{CSST Collaboration}}},
  \bibinfo {author} {\bibfnamefont {Y.}~\bibnamefont {{Gong}}}, \bibinfo
  {author} {\bibfnamefont {H.}~\bibnamefont {{Miao}}}, \bibinfo {author}
  {\bibfnamefont {H.}~\bibnamefont {{Zhan}}}, \bibinfo {author} {\bibfnamefont
  {Z.-Y.}\ \bibnamefont {{Li}}}, \bibinfo {author} {\bibfnamefont
  {J.}~\bibnamefont {{Shangguan}}}, \bibinfo {author} {\bibfnamefont
  {H.}~\bibnamefont {{Li}}}, \bibinfo {author} {\bibfnamefont {C.}~\bibnamefont
  {{Liu}}}, \bibinfo {author} {\bibfnamefont {X.}~\bibnamefont {{Chen}}},
  \bibinfo {author} {\bibfnamefont {H.}~\bibnamefont {{Yuan}}},  \emph
  {et~al.},\ }\bibfield  {title} {\enquote {\bibinfo {title} {{Introduction to
  the China Space Station Telescope (CSST)}},}\ }\href {\doibase
  10.48550/arXiv.2507.04618} {\bibfield  {journal} {\bibinfo  {journal} {arXiv
  e-prints}\ ,\ \bibinfo {eid} {arXiv:2507.04618}} (\bibinfo {year} {2025})},\
  \Eprint {http://arxiv.org/abs/2507.04618}{arXiv:2507.04618
  [astro-ph.IM]}\BibitemShut {NoStop}%
\bibitem [{\citenamefont {{Villaescusa-Navarro}}\ \emph
  {et~al.}(2020)\citenamefont {{Villaescusa-Navarro}}, \citenamefont {{Hahn}},
  \citenamefont {{Massara}}, \citenamefont {{Banerjee}}, \citenamefont
  {{Delgado}}, \citenamefont {{Ramanah}}, \citenamefont {{Charnock}},
  \citenamefont {{Giusarma}}, \citenamefont {{Li}}, \citenamefont {{Allys}}
  \emph {et~al.}}]{2020ApJS..250....2V}%
  \BibitemOpen
  \bibfield  {author} {\bibinfo {author} {\bibfnamefont {F.}~\bibnamefont
  {{Villaescusa-Navarro}}}, \bibinfo {author} {\bibfnamefont {C.}~\bibnamefont
  {{Hahn}}}, \bibinfo {author} {\bibfnamefont {E.}~\bibnamefont {{Massara}}},
  \bibinfo {author} {\bibfnamefont {A.}~\bibnamefont {{Banerjee}}}, \bibinfo
  {author} {\bibfnamefont {A.~M.}\ \bibnamefont {{Delgado}}}, \bibinfo {author}
  {\bibfnamefont {D.~K.}\ \bibnamefont {{Ramanah}}}, \bibinfo {author}
  {\bibfnamefont {T.}~\bibnamefont {{Charnock}}}, \bibinfo {author}
  {\bibfnamefont {E.}~\bibnamefont {{Giusarma}}}, \bibinfo {author}
  {\bibfnamefont {Y.}~\bibnamefont {{Li}}}, \bibinfo {author} {\bibfnamefont
  {E.}~\bibnamefont {{Allys}}},  \emph {et~al.},\ }\bibfield  {title} {\enquote
  {\bibinfo {title} {{The Quijote Simulations}},}\ }\href {\doibase
  10.3847/1538-4365/ab9d82} {\bibfield  {journal} {\bibinfo  {journal} {\apjs}\
  }\textbf {\bibinfo {volume} {250}},\ \bibinfo {eid} {2} (\bibinfo {year}
  {2020})},\ \Eprint {http://arxiv.org/abs/1909.05273}{arXiv:1909.05273
  [astro-ph.CO]}\BibitemShut {NoStop}%
\bibitem [{\citenamefont {{Chuang}}\ \emph {et~al.}(2019)\citenamefont
  {{Chuang}}, \citenamefont {{Yepes}}, \citenamefont {{Kitaura}}, \citenamefont
  {{Pellejero-Ibanez}}, \citenamefont {{Rodr{\'\i}guez-Torres}}, \citenamefont
  {{Feng}}, \citenamefont {{Metcalf}}, \citenamefont {{Wechsler}},
  \citenamefont {{Zhao}}, \citenamefont {{To}} \emph
  {et~al.}}]{2019MNRAS.487...48C}%
  \BibitemOpen
  \bibfield  {author} {\bibinfo {author} {\bibfnamefont {C.-H.}\ \bibnamefont
  {{Chuang}}}, \bibinfo {author} {\bibfnamefont {G.}~\bibnamefont {{Yepes}}},
  \bibinfo {author} {\bibfnamefont {F.-S.}\ \bibnamefont {{Kitaura}}}, \bibinfo
  {author} {\bibfnamefont {M.}~\bibnamefont {{Pellejero-Ibanez}}}, \bibinfo
  {author} {\bibfnamefont {S.}~\bibnamefont {{Rodr{\'\i}guez-Torres}}},
  \bibinfo {author} {\bibfnamefont {Y.}~\bibnamefont {{Feng}}}, \bibinfo
  {author} {\bibfnamefont {R.~B.}\ \bibnamefont {{Metcalf}}}, \bibinfo {author}
  {\bibfnamefont {R.~H.}\ \bibnamefont {{Wechsler}}}, \bibinfo {author}
  {\bibfnamefont {C.}~\bibnamefont {{Zhao}}}, \bibinfo {author} {\bibfnamefont
  {C.-H.}\ \bibnamefont {{To}}},  \emph {et~al.},\ }\bibfield  {title}
  {\enquote {\bibinfo {title} {{UNIT project: Universe N-body simulations for
  the Investigation of Theoretical models from galaxy surveys}},}\ }\href
  {\doibase 10.1093/mnras/stz1233} {\bibfield  {journal} {\bibinfo  {journal}
  {\mnras}\ }\textbf {\bibinfo {volume} {487}},\ \bibinfo {pages} {48}
  (\bibinfo {year} {2019})},\ \Eprint
  {http://arxiv.org/abs/1811.02111}{arXiv:1811.02111 [astro-ph.CO]}\BibitemShut
  {NoStop}%
\bibitem [{\citenamefont {{Cautun}}\ \emph {et~al.}(2018)\citenamefont
  {{Cautun}}, \citenamefont {{Paillas}}, \citenamefont {{Cai}}, \citenamefont
  {{Bose}}, \citenamefont {{Armijo}}, \citenamefont {{Li}},\ and\ \citenamefont
  {{Padilla}}}]{2018MNRAS.476.3195C}%
  \BibitemOpen
  \bibfield  {author} {\bibinfo {author} {\bibfnamefont {M.}~\bibnamefont
  {{Cautun}}}, \bibinfo {author} {\bibfnamefont {E.}~\bibnamefont {{Paillas}}},
  \bibinfo {author} {\bibfnamefont {Y.-C.}\ \bibnamefont {{Cai}}}, \bibinfo
  {author} {\bibfnamefont {S.}~\bibnamefont {{Bose}}}, \bibinfo {author}
  {\bibfnamefont {J.}~\bibnamefont {{Armijo}}}, \bibinfo {author}
  {\bibfnamefont {B.}~\bibnamefont {{Li}}}, \ and\ \bibinfo {author}
  {\bibfnamefont {N.}~\bibnamefont {{Padilla}}},\ }\bibfield  {title} {\enquote
  {\bibinfo {title} {{The Santiago-Harvard-Edinburgh-Durham void comparison -
  I. SHEDding light on chameleon gravity tests}},}\ }\href {\doibase
  10.1093/mnras/sty463} {\bibfield  {journal} {\bibinfo  {journal} {\mnras}\
  }\textbf {\bibinfo {volume} {476}},\ \bibinfo {pages} {3195} (\bibinfo {year}
  {2018})},\ \Eprint {http://arxiv.org/abs/1710.01730}{arXiv:1710.01730
  [astro-ph.CO]}\BibitemShut {NoStop}%
\bibitem [{\citenamefont {{Prada}}\ \emph {et~al.}(2012)\citenamefont
  {{Prada}}, \citenamefont {{Klypin}}, \citenamefont {{Cuesta}}, \citenamefont
  {{Betancort-Rijo}},\ and\ \citenamefont {{Primack}}}]{2012MNRAS.423.3018P}%
  \BibitemOpen
  \bibfield  {author} {\bibinfo {author} {\bibfnamefont {F.}~\bibnamefont
  {{Prada}}}, \bibinfo {author} {\bibfnamefont {A.~A.}\ \bibnamefont
  {{Klypin}}}, \bibinfo {author} {\bibfnamefont {A.~J.}\ \bibnamefont
  {{Cuesta}}}, \bibinfo {author} {\bibfnamefont {J.~E.}\ \bibnamefont
  {{Betancort-Rijo}}}, \ and\ \bibinfo {author} {\bibfnamefont
  {J.}~\bibnamefont {{Primack}}},\ }\bibfield  {title} {\enquote {\bibinfo
  {title} {{Halo concentrations in the standard {\ensuremath{\Lambda}} cold
  dark matter cosmology}},}\ }\href {\doibase 10.1111/j.1365-2966.2012.21007.x}
  {\bibfield  {journal} {\bibinfo  {journal} {\mnras}\ }\textbf {\bibinfo
  {volume} {423}},\ \bibinfo {pages} {3018} (\bibinfo {year} {2012})},\ \Eprint
  {http://arxiv.org/abs/1104.5130}{arXiv:1104.5130 [astro-ph.CO]}\BibitemShut
  {NoStop}%
\bibitem [{\citenamefont {{Feng}}\ \emph {et~al.}(2016)\citenamefont {{Feng}},
  \citenamefont {{Chu}}, \citenamefont {{Seljak}},\ and\ \citenamefont
  {{McDonald}}}]{2016MNRAS.463.2273F}%
  \BibitemOpen
  \bibfield  {author} {\bibinfo {author} {\bibfnamefont {Y.}~\bibnamefont
  {{Feng}}}, \bibinfo {author} {\bibfnamefont {M.-Y.}\ \bibnamefont {{Chu}}},
  \bibinfo {author} {\bibfnamefont {U.}~\bibnamefont {{Seljak}}}, \ and\
  \bibinfo {author} {\bibfnamefont {P.}~\bibnamefont {{McDonald}}},\ }\bibfield
   {title} {\enquote {\bibinfo {title} {{FASTPM: a new scheme for fast
  simulations of dark matter and haloes}},}\ }\href {\doibase
  10.1093/mnras/stw2123} {\bibfield  {journal} {\bibinfo  {journal} {\mnras}\
  }\textbf {\bibinfo {volume} {463}},\ \bibinfo {pages} {2273} (\bibinfo {year}
  {2016})},\ \Eprint {http://arxiv.org/abs/1603.00476}{arXiv:1603.00476
  [astro-ph.CO]}\BibitemShut {NoStop}%
\bibitem [{\citenamefont {{Planck Collaboration}}\ \emph
  {et~al.}(2016)\citenamefont {{Planck Collaboration}}, \citenamefont {{Ade}},
  \citenamefont {{Aghanim}}, \citenamefont {{Arnaud}}, \citenamefont
  {{Ashdown}}, \citenamefont {{Aumont}}, \citenamefont {{Baccigalupi}},
  \citenamefont {{Banday}}, \citenamefont {{Barreiro}}, \citenamefont
  {{Bartlett}} \emph {et~al.}}]{2016A&A...594A..13P}%
  \BibitemOpen
  \bibfield  {author} {\bibinfo {author} {\bibnamefont {{Planck
  Collaboration}}}, \bibinfo {author} {\bibfnamefont {P.~A.~R.}\ \bibnamefont
  {{Ade}}}, \bibinfo {author} {\bibfnamefont {N.}~\bibnamefont {{Aghanim}}},
  \bibinfo {author} {\bibfnamefont {M.}~\bibnamefont {{Arnaud}}}, \bibinfo
  {author} {\bibfnamefont {M.}~\bibnamefont {{Ashdown}}}, \bibinfo {author}
  {\bibfnamefont {J.}~\bibnamefont {{Aumont}}}, \bibinfo {author}
  {\bibfnamefont {C.}~\bibnamefont {{Baccigalupi}}}, \bibinfo {author}
  {\bibfnamefont {A.~J.}\ \bibnamefont {{Banday}}}, \bibinfo {author}
  {\bibfnamefont {R.~B.}\ \bibnamefont {{Barreiro}}}, \bibinfo {author}
  {\bibfnamefont {J.~G.}\ \bibnamefont {{Bartlett}}},  \emph {et~al.},\
  }\bibfield  {title} {\enquote {\bibinfo {title} {{Planck 2015 results. XIII.
  Cosmological parameters}},}\ }\href {\doibase 10.1051/0004-6361/201525830}
  {\bibfield  {journal} {\bibinfo  {journal} {\aap}\ }\textbf {\bibinfo
  {volume} {594}},\ \bibinfo {eid} {A13} (\bibinfo {year} {2016})},\ \Eprint
  {http://arxiv.org/abs/1502.01589}{arXiv:1502.01589 [astro-ph.CO]}\BibitemShut
  {NoStop}%
\bibitem [{\citenamefont {{Komatsu}}\ \emph {et~al.}(2009)\citenamefont
  {{Komatsu}}, \citenamefont {{Dunkley}}, \citenamefont {{Nolta}},
  \citenamefont {{Bennett}}, \citenamefont {{Gold}}, \citenamefont {{Hinshaw}},
  \citenamefont {{Jarosik}}, \citenamefont {{Larson}}, \citenamefont {{Limon}},
  \citenamefont {{Page}} \emph {et~al.}}]{2009ApJS..180..330K}%
  \BibitemOpen
  \bibfield  {author} {\bibinfo {author} {\bibfnamefont {E.}~\bibnamefont
  {{Komatsu}}}, \bibinfo {author} {\bibfnamefont {J.}~\bibnamefont
  {{Dunkley}}}, \bibinfo {author} {\bibfnamefont {M.~R.}\ \bibnamefont
  {{Nolta}}}, \bibinfo {author} {\bibfnamefont {C.~L.}\ \bibnamefont
  {{Bennett}}}, \bibinfo {author} {\bibfnamefont {B.}~\bibnamefont {{Gold}}},
  \bibinfo {author} {\bibfnamefont {G.}~\bibnamefont {{Hinshaw}}}, \bibinfo
  {author} {\bibfnamefont {N.}~\bibnamefont {{Jarosik}}}, \bibinfo {author}
  {\bibfnamefont {D.}~\bibnamefont {{Larson}}}, \bibinfo {author}
  {\bibfnamefont {M.}~\bibnamefont {{Limon}}}, \bibinfo {author} {\bibfnamefont
  {L.}~\bibnamefont {{Page}}},  \emph {et~al.},\ }\bibfield  {title} {\enquote
  {\bibinfo {title} {{Five-Year Wilkinson Microwave Anisotropy Probe
  Observations: Cosmological Interpretation}},}\ }\href {\doibase
  10.1088/0067-0049/180/2/330} {\bibfield  {journal} {\bibinfo  {journal}
  {\apjs}\ }\textbf {\bibinfo {volume} {180}},\ \bibinfo {pages} {330}
  (\bibinfo {year} {2009})},\ \Eprint
  {http://arxiv.org/abs/0803.0547}{arXiv:0803.0547 [astro-ph]}\BibitemShut
  {NoStop}%
\bibitem [{\citenamefont {{Klypin}}\ \emph {et~al.}(2011)\citenamefont
  {{Klypin}}, \citenamefont {{Trujillo-Gomez}},\ and\ \citenamefont
  {{Primack}}}]{2011ApJ...740..102K}%
  \BibitemOpen
  \bibfield  {author} {\bibinfo {author} {\bibfnamefont {A.~A.}\ \bibnamefont
  {{Klypin}}}, \bibinfo {author} {\bibfnamefont {S.}~\bibnamefont
  {{Trujillo-Gomez}}}, \ and\ \bibinfo {author} {\bibfnamefont
  {J.}~\bibnamefont {{Primack}}},\ }\bibfield  {title} {\enquote {\bibinfo
  {title} {{Dark Matter Halos in the Standard Cosmological Model: Results from
  the Bolshoi Simulation}},}\ }\href {\doibase 10.1088/0004-637X/740/2/102}
  {\bibfield  {journal} {\bibinfo  {journal} {\apj}\ }\textbf {\bibinfo
  {volume} {740}},\ \bibinfo {eid} {102} (\bibinfo {year} {2011})},\ \Eprint
  {http://arxiv.org/abs/1002.3660}{arXiv:1002.3660 [astro-ph.CO]}\BibitemShut
  {NoStop}%
\bibitem [{\citenamefont {{Li}}\ \emph {et~al.}(2013)\citenamefont {{Li}},
  \citenamefont {{Zhao}},\ and\ \citenamefont {{Koyama}}}]{ECOSMOG_V_1}%
  \BibitemOpen
  \bibfield  {author} {\bibinfo {author} {\bibfnamefont {B.}~\bibnamefont
  {{Li}}}, \bibinfo {author} {\bibfnamefont {G.-B.}\ \bibnamefont {{Zhao}}}, \
  and\ \bibinfo {author} {\bibfnamefont {K.}~\bibnamefont {{Koyama}}},\
  }\bibfield  {title} {\enquote {\bibinfo {title} {{Exploring Vainshtein
  mechanism on adaptively refined meshes}},}\ }\href {\doibase
  10.1088/1475-7516/2013/05/023} {\bibfield  {journal} {\bibinfo  {journal}
  {\jcap}\ }\textbf {\bibinfo {volume} {2013}},\ \bibinfo {eid} {023} (\bibinfo
  {year} {2013})},\ \Eprint {http://arxiv.org/abs/1303.0008}{arXiv:1303.0008
  [astro-ph.CO]}\BibitemShut {NoStop}%
\bibitem [{\citenamefont {{Hinshaw}}\ \emph {et~al.}(2013)\citenamefont
  {{Hinshaw}}, \citenamefont {{Larson}}, \citenamefont {{Komatsu}},
  \citenamefont {{Spergel}}, \citenamefont {{Bennett}}, \citenamefont
  {{Dunkley}}, \citenamefont {{Nolta}}, \citenamefont {{Halpern}},
  \citenamefont {{Hill}}, \citenamefont {{Odegard}} \emph
  {et~al.}}]{2013ApJS..208...19H}%
  \BibitemOpen
  \bibfield  {author} {\bibinfo {author} {\bibfnamefont {G.}~\bibnamefont
  {{Hinshaw}}}, \bibinfo {author} {\bibfnamefont {D.}~\bibnamefont {{Larson}}},
  \bibinfo {author} {\bibfnamefont {E.}~\bibnamefont {{Komatsu}}}, \bibinfo
  {author} {\bibfnamefont {D.~N.}\ \bibnamefont {{Spergel}}}, \bibinfo {author}
  {\bibfnamefont {C.~L.}\ \bibnamefont {{Bennett}}}, \bibinfo {author}
  {\bibfnamefont {J.}~\bibnamefont {{Dunkley}}}, \bibinfo {author}
  {\bibfnamefont {M.~R.}\ \bibnamefont {{Nolta}}}, \bibinfo {author}
  {\bibfnamefont {M.}~\bibnamefont {{Halpern}}}, \bibinfo {author}
  {\bibfnamefont {R.~S.}\ \bibnamefont {{Hill}}}, \bibinfo {author}
  {\bibfnamefont {N.}~\bibnamefont {{Odegard}}},  \emph {et~al.},\ }\bibfield
  {title} {\enquote {\bibinfo {title} {{Nine-year Wilkinson Microwave
  Anisotropy Probe (WMAP) Observations: Cosmological Parameter Results}},}\
  }\href {\doibase 10.1088/0067-0049/208/2/19} {\bibfield  {journal} {\bibinfo
  {journal} {\apjs}\ }\textbf {\bibinfo {volume} {208}},\ \bibinfo {eid} {19}
  (\bibinfo {year} {2013})},\ \Eprint
  {http://arxiv.org/abs/1212.5226}{arXiv:1212.5226 [astro-ph.CO]}\BibitemShut
  {NoStop}%
\end{thebibliography}%

\end{document}